\documentclass[aps,prx,twocolumn,superscriptaddress,export,nofootinbib]{revtex4-1}
\usepackage[colorlinks=true,linkcolor=blue,citecolor=blue,urlcolor=blue]{hyperref}
\usepackage[utf8]{inputenc}
\usepackage[german, english]{babel}
\usepackage[dvipsnames]{xcolor}
\usepackage{amsmath,amsfonts, amssymb, amsthm, dsfont}
\usepackage{bm}
\usepackage{graphicx}
\usepackage{tikz}
\usepackage{physics}
\usepackage{xspace}
\usepackage[export]{adjustbox}

\usepackage{enumerate}
\usepackage{subcaption}
\usepackage[justification=raggedright,format=plain]{caption}
\usepackage{bbold}
\usepackage[normalem]{ulem}

\newcommand{\ghz}{\mathrm{GHZ}}
\newcommand{\bmm}{{\bm{m}}}
\newcommand{\bmh}{{\bm{h}}}

\newcommand{\gammatilde}{\widetilde{\gamma}}
\newcommand{\ba}{\mathbf{a}}

\newcommand{\calZ}{\mathcal{Z}}

\newcommand{\calN}{\mathcal{N}}

\newcommand{\RBIM}{\text{RBIM}}
\newcommand{\Ising}{\text{Ising}}
\newcommand{\bzero}{\mathbf{0}}

\newcommand{\blambda}{{\bm{\lambda}}}

\begin{document}

\title{The Stability of Gapped Quantum Matter and\\ Error-Correction with Adiabatic Noise}
\author{Ali Lavasani}
\affiliation{Kavli Institute for Theoretical Physics, Santa Barbara, CA 93106, USA}
				
\author{Sagar Vijay}
\affiliation{Department of Physics, University of California, Santa Barbara, CA 93106, USA}			

\begin{abstract}
The codespace of a quantum error-correcting code can often be identified with the degenerate ground-space within a gapped phase of quantum matter.
We argue that the stability of such a phase is directly related to a set of coherent error processes against which this quantum error-correcting code (QECC) is robust: such a quantum code can recover from \emph{adiabatic noise channels}, corresponding to random adiabatic drift of code states through the phase, with asymptotically perfect fidelity in the thermodynamic limit, as long as this adiabatic evolution keeps states sufficiently ``close" to the initial ground-space.  We further argue that when specific decoders -- such as minimum-weight perfect matching -- are applied to recover this information, an error-correcting threshold is generically encountered \emph{within} the gapped phase.  In cases where the adiabatic evolution is known, we explicitly show examples in which quantum information can be recovered by using stabilizer measurements and Pauli feedback, even up to a phase boundary, though the resulting decoding transitions are in different universality classes from the optimal decoding transitions in the presence of incoherent Pauli noise. This provides  examples where non-local, coherent noise effectively decoheres in the presence of syndrome measurements in a stabilizer QECC. 
\end{abstract}

 \maketitle

 Quantum error correcting codes (QECC) use entangled quantum states as basis states for qubits, so that sufficiently weak coupling to an external environment cannot corrupt this encoded information.  The codespace for a QECC, defined as the Hilbert space spanned by the entangled basis states, can often be identified with the degenerate ground-space of a gapped quantum many-body Hamiltonian. This Hamiltonian is generally observed to be robust to a class of weak local perturbations, so that the ground-space is representative of a stable \emph{zero-temperature phase} of quantum matter. As an example, the codespace for the quantum repetition code can be exactly identified with the ground-space of a $d$-dimensional quantum Ising model in the absence of a transverse field. This ground-space describes a point within the symmetry-broken phase of the quantum Ising model, and exhibits key properties (e.g. two-fold degeneracy, and ferromagnetic, long-range order) which are universal, and persist throughout the phase.  
 
The relationship between the codespace of a QECC and gapped phases of quantum matter remains to be fully explored.  For topological QECC's, some progress has been made to elucidate this connection.  It has been rigorously shown \cite{bravyi2010topological,bravyi2011short,michalakis2013stability} that if the codespace $\mathcal{C}$ of a topological code satisfies additional conditions -- if $\mathcal{C}$ has a macroscopically-large code distance, and local and global projectors onto $\mathcal{C}$ are sufficiently similar -- then a  Hamiltonian for which $\mathcal{C}$ is its ground-space, is stable to weak, local perturbations, and thus describes a gapped phase of quantum  matter.  All models of two-dimensional topological quantum orders formed of frustration-free, commuting projector Hamiltonians (e.g. Levin-Wen string-net models \cite{levin2005string}) satisfy these conditions \cite{michalakis2013stability}.

\begin{figure}
    \includegraphics[width=0.81\columnwidth]{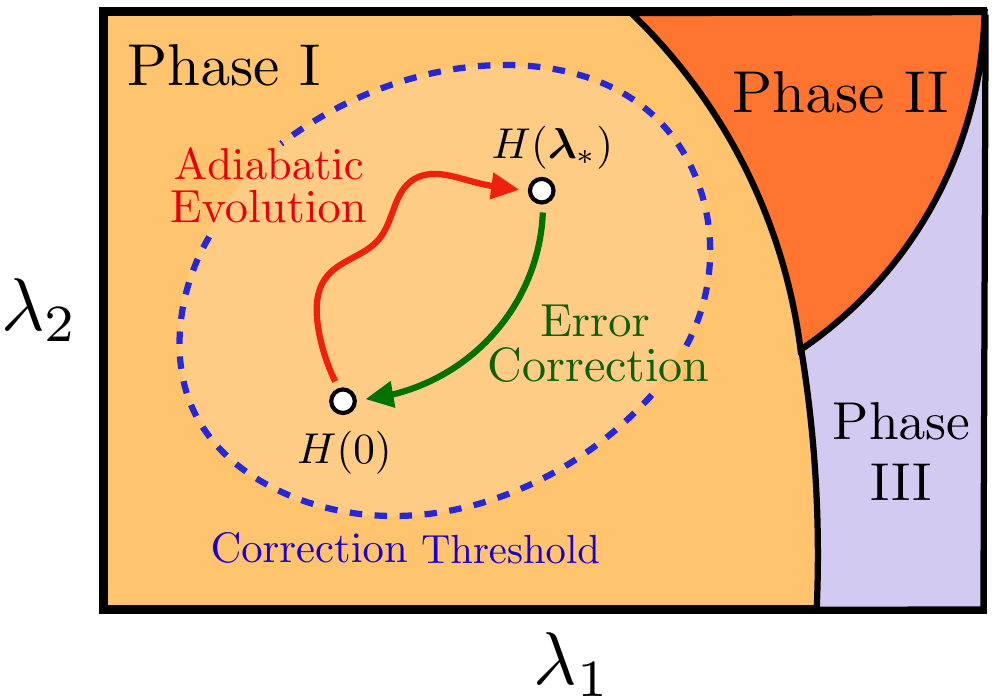}
    \caption{Quantum information is encoded in the ground-space of $H(0)$, a quantum many-body Hamiltonian within a gapped quantum phase. We argue that stability of this phase implies that after adiabatic evolution within a finite neighborhood of $H(0)$, it is still possible to recover the encoded information.  An error-correction threshold can be encountered within the gapped phase, outside of which the recovery of quantum information is impossible. 
     }
    \label{fig_schematic}
\end{figure}

In this work, we introduce a new connection between the stability of a gapped quantum phase and the properties of a QECC formed from its ground-space.    Consider a quantum many-body Hamiltonian whose degenerate ground-space describes a gapped phase of quantum matter, and is used to encode quantum information.  A state is prepared within this ground-space and subsequently evolves adiabatically within the same gapped phase.  When this adiabatic drift of the codespace is \emph{a priori} unknown, we argue that it remains possible to recover the encoded information with asymptotically perfect fidelity in the thermodynamic limit as long as the adiabatic evolution remains sufficiently close to the original gapped ground-space.  We refer to this unknown adiabatic evolution as an \emph{adiabatic noise channel}.  We show this statement to be true for zero-form symmetry-protected topological phases, and symmetry-breaking, gapped phases by introducing a generalization of the Knill-Laflamme conditions \cite{knill1997theory} for adiabatic error channels, and argue this to be the case for topological QECC's such as the toric code.   When specific decoding schemes -- such as minimum-weight perfect matching -- are applied to recover this quantum information, we further argue that there will be an error-correction threshold \emph{within} the gapped phase, so that if the adiabatic drift takes the code state sufficiently far away from the original ground-space,  then quantum information recovery will not be possible. Second, when the adiabatic evolution of the code-state is known, then syndrome measurements and Pauli feedback can be used, in specific examples that we consider, to successfully decode the encoded information, even up to a phase boundary.  Results ($i$) and ($ii$) are obtained through a combination of exact calculations and numerical studies of the repetition code and the toric code. 

Our work relates to recent efforts to connect quantum dynamics and quantum information recovery.  Notably, Refs. \cite{bao2023mixed,fan2023diagnostics,lee2023quantum, chen2023separability, su2024tapestry} have shown that the ability to recover quantum information encoded in a topological QECC, after the application of a local error channel $\mathcal{N}$, is directly related to a phase transition in the intrinsic topological order characterizing the density matrix $\mathcal{N}(\ketbra{\psi_0})$. The adiabatic noise channels considered in our work are coherent, and intrinsically non-local, thus providing an alternate setting in which error-correction remains possible.  
Other recent work \cite{lake2022exact} has designed unitary circuits to recognize certain symmetry-protected topological (SPT) orders, though the question of whether logical information encoded within the ground-space is preserved during this process has not been considered. 

{The settings  we consider in this work arise naturally in the context of passive quantum error correction, where quantum  information is encoded in a codespace and subsequently evolves according to a Hamiltonian, though the ground-space of the latter is not necessarily the codespace of interest. Previous studies on  this subject have mostly focused on thermal stability of quantum memories by coupling the codespace to a thermal bath \cite{alicki2009thermalization,landon2013local,brown2016quantum} or the lifetime of quantum information in specific quantum error-correcting codes evolving under a sudden quench \cite{pastawski2009limitations,brell2014thermalization,mazza2013robustness,pastawski2012quantum}. In contrast, we assume here that the perturbation is turned on adiabatically and study the possibility of reversing it via recovery maps of the underlying QECC. The applicability of our results to more general models of passive error-correction, e.g. that keep the final state within the same gapped phase, provides an open direction of inquiry.}

We emphasize that adiabatic drift inside a gapped phase is an instance of coherent noise affecting a QECC, which is an active field of study (see~\cite{greenbaum2017modeling,beale2018quantum,huang2019performance,gutierrez2016errors,darmawan2017tensor,suzuki2017efficient} for an incomplete list of references). In particular, in Ref.~\cite{beale2018quantum} it was shown that measuring the stabilizers of a stabilizer code decoheres \emph{local} coherent errors, meaning that the noise model plus stabilizer measurements would effectively converge to a probabilistic, local Pauli error model.  
Our results suggest that stabilizer measurements can effectively decohere certain kinds of adiabatic noise in the thermodynamic limit, even though the latter is a highly non-local, coherent process. How local measurements can decohere a non-local, quantum-coherent process, and the resulting Pauli error models that emerge, remain open questions of study.

  \tableofcontents

  \section{Setup and Summary of Results}\label{sec_general_setup}
  We now formally define the general setup that we will consider in this work. Let $\mathcal{C}$ denote a quantum stabilizer code encoding $k$ logical qubits in $N$ physical qubits and let $\{g_1,\cdots,g_M\}$ be a complete set of generators for its stabilizer group. Let $\ket{\varphi_i}$ for $i=1,\cdots,2^k$, be an orthonormal basis for the codespace of $\mathcal{C}$. Define the corresponding code Hamiltonian as $H_\mathcal{C}=-\sum_{m=1}^M g_m$, whose $2^k$-fold degenerate ground state subspace coincides with the codespace of $\mathcal{C}$. Let $V(\blambda)=\sum_{j=1}^N V_j(\blambda)$ with $\blambda \in \mathbb{R}^n$, denote a continuous family of local perturbations; formally, each $V_j(\blambda)$ is a smooth function of $\blambda$ and its norm (as well as the norms of $\partial_\blambda V_j$ and $\partial^2_\blambda V_j$) is bounded by a constant $J$, and is localized around qubit $j$, with $V_j(\bzero)=0$. Let $H(\blambda)=H_\mathcal{C}+V(\blambda)$. Let $E_i(\blambda)$ denote the $i$'th lowest energy  of $H(\blambda)$. We assume that there is a neighborhood $\mathcal{A}\subset \mathbb{R}^n$  around $\bzero$, for which $H(\blambda)$ remains gapped in the thermodynamic limit, with a minimum energy gap $\Delta$\footnote{Formally, this means that $\Delta\equiv\min_{\blambda\in\mathcal{A}} (E_{2^k+1}-E_{2^k})$ remains bounded from below by a non-zero value in the thermodynamic limit, while $\delta=\max_{\blambda\in \mathcal{A}}(E_{2^k}-E_1)$ goes to zero in the thermodynamic limit faster than any polynomial in $N$. }.

Let $\ket{\psi}=\sum_{i=1}^{2^k} \alpha_{i}\ket{\varphi_i}$ denote a code state, encoded in the ground state subspace of $H(\bzero)$. We now imagine tuning the parameter $\blambda$ adiabatically along a path $\gamma(t)\in \mathcal{A}$ from $\gamma(0)=\bzero$ to a finite value $\gamma(T)=\blambda_{*}$. Formally, this requires that $T\gg {J^2\,N^2}/{\Delta^3}$ to ensure the system always remains in the approximate ground state subspace \cite{albash2018adiabatic,jansen2007bounds} while $T\, \delta \ll 1$, where $\delta$ is the finite-size splitting between ground-states, so that the dynamical phase difference accumulated during the adiabatic process  is  negligible \cite{nayak2008non}. We denote the resulting unitary evolution operator as $\mathcal{U}_\gamma$. We define an adiabatic noise channel over $\mathcal{A}$ as 
\begin{align}
\mathcal{E}_\Gamma(\rho)=\sum_{\gamma\in\Gamma} p_\gamma~\mathcal{U}_{\gamma}\,\rho\, \mathcal{U}_{\gamma}^\dagger    
\end{align}
where $\Gamma$ is a set of adiabatic paths in $\mathcal{A}$ and $p_\gamma$ is a probability distribution over $\Gamma$.

In this work we investigate the validity of the following two statements:
\begin{itemize}
    \item \textbf{Statement I:} If provided with the state $\ket{\widetilde{\psi}}\equiv\mathcal{U}_\gamma\ket{\psi}$ and the knowledge of the adiabatic path $\gamma$, one can recover $\ket{\psi}$ with perfect fidelity in the thermodynamic limit by measuring the operators $\{g_1,\cdots,g_M\}$ and applying Pauli feedback, {which can depend on $\gamma$ as well as the measurement outcomes}.
    \item \textbf{Statement II:} If provided with the state $\widetilde{\rho}=\mathcal{E}_\Gamma(\ketbra{\psi})$, one can  recover $\ket{\psi}$ with perfect fidelity in the thermodynamic limit. 
\end{itemize}

These statements formalize certain natural connections that can be drawn between the stability of gapped phases of quantum matter and quantum error-correction.   Statement I makes precise the notion that one may move the ground-state of a gapped phase back to a ``fixed-point" by measuring the excitations of the fixed point Hamiltonian and removing these via appropriate feedback operations.  The scenario considered in Statement I relies on knowledge of the adiabatic path $\gamma$, and restricts the recovery channel to involve stabilizer measurements and Pauli feedback.  In contrast, Statement II is intended to formalize a potentially practically-relevant aspect of the relation between a gapped phase of matter and  error-correction, by studying whether it is possible to recover quantum information from \emph{unknown} evolution within the  phase. We also do not limit the recovery channel in Statement II to stabilizer measurements and Pauli feedback, to study the broad validity of this statement\footnote{Without restrictions on the recovery channel in Statement I, one could trivially recover $\ket{\psi}$ by applying $\mathcal{U}_\gamma^\dagger$. Analogously, Statement II is trivially true if $\Gamma$ contains only one element.}.

We emphasize that Statements I and II are also analogous to situations that are naturally considered in the study of quantum error-correction in the presence of \emph{incoherent} noise.  In that setting, knowledge of the incoherent noise which has affected the codespace (e.g. a Pauli noise channel with a known noise rate) will necessarily be used by an optimal decoder to successfully recover quantum information when the noise rate is sufficiently low. 
The threshold corresponding to the noise rate below which optimal decoding is successful is generally reduced when the rate of incoherent noise is unknown. Knowledge of the error rate is analogous, in our problem, to knowledge of the adiabatic path $\gamma$ along which the codespace has evolved in Statement I. Thus, an analogous result may be expected (as we subsequently argue), that even when Pauli recovery is possible for each path $\gamma \in \Gamma$, it is not necessarily possible to recover quantum information after the noise channel $\mathcal{E}_{\Gamma}$.

The remainder of our work is organized as follows.  Sections \ref{sec_rev_adi_evo} and \ref{sec_adiabatic_channel} are dedicated to investigating the validity of Statements I and II, respectively. In Sec. \ref{sec_rev_adi_evo}, we study  examples involving the quantum repetition code, one-dimensional symmetry-protected topological (SPT) orders, and the toric code, in which Pauli feedback can be used to recover from known adiabatic evolution within the corresponding gapped phases.  We demonstrate that quantum error-correction, with asymptotically perfect fidelity in the thermodynamic limit, remains possible for certain adiabatic paths within these phases, and all the way up to a critical point that separates these from other phases of matter.  Intriguingly for the toric code, the decoding fidelity sees a phase transition in the two-dimensional Ising universality class; this is in stark contrast to the two-dimensional random bond Ising universality class for the optimal decoding transition with perfect stabilizer measurements and incoherent Pauli noise. 

An important implication of these results is that syndrome measurements in these stabilizer QECC's can completely decohere the coherent noise corresponding to adiabatic evolution in the thermodynamic limit.  The state of the system after syndrome measurements can thus be obtained by an effective \emph{incoherent} Pauli noise model on the initial state, where the noise remains short-range correlated when the adiabatic evolution lies close to the fixed-point.  These results support the conjecture that Statement I holds for known adiabatic paths that lie sufficiently close to the fixed-point of any gapped phase, though the validity of this statement for paths $\gamma$ approaching a phase boundary is unclear.

{In Sec. \ref{sec_adiabatic_channel}, we study error-correction in the presence of an adiabatic noise channel.  We provide an explicit example of an adiabatic noise channel for the quantum repetition code, which keeps the codespace within the gapped phase, in which error correction is nonetheless impossible. We further argue that decoders for the toric code which do not rely on specific knowledge of the noise model (e.g. minimum-weight perfect-matching) will fail for an adiabatic noise channel that takes the codespace outside of a finite neighborhood of the fixed-point that nevertheless remains entirely within the gapped phase.  To supplement these examples, we derive  conditions for the existence of a recovery map after an adiabatic noise channel is applied to a general gapped phase \cite{nielsen2002quantum}; intuitively, these conditions guarantee the successful recovery of quantum information when the ground-spaces in different parts of the gapped phase diagram are sufficiently ``similar".  We argue that these conditions are generally fulfilled for symmetry-broken phases of matter, and zero-form symmetry-protected topological phases when the adiabatic path remains sufficiently ``close" to the initial state.  These results provide evidence that Statement II holds for any gapped quantum phase, within a finite neighborhood of the fixed-point. }


\section{Reversing  Adiabatic Evolution with Stabilizer Measurements and  Feedback}\label{sec_rev_adi_evo}
In this section we will consider two examples where one can reverse adiabatic evolution, as specified by a path $\gamma$, via stabilizer measurements and Pauli feedback, as long as $\gamma$ remains entirely within the gapped phase. The first example is adiabatic evolution within the ferromagnetic phase of the one-dimensional transverse field Ising chain (TFIC) whose ``fixed-point", in the absence of a transverse field, corresponds to the Hamiltonian whose ground-space describes the quantum repetition code. The second example we consider is the two-dimensional toric code in the presence of local perturbations. 

In both examples, we show that as long as $\gamma$ stays inside the gapped phase, one can return to the codespace via stabilizer measurements and by applying proper feedback. Interestingly, the decoder in the TFIC example does not use the knowledge of the adiabatic path $\gamma$, and the decoder for the perturbed toric code, uses only $\gamma(T)$, i.e. the end point of the adiabatic evolution. 

\subsection{Transverse Field Ising Chain (TFIC)}\label{sec_repetition_code_and_1dTFIM}

Consider a chain of $N$ qubits with closed boundary conditions, and described by the following Hamiltonian:
\begin{align}\label{eq:orginial_hamiltonina}
    H_g=- \sum_{i=1}^N Z_i Z_{i+1} - g \sum_{i=1}^N X_i.
\end{align}
Let $P=\prod_{i=1}^N X_i$ denote the Ising symmetry transformation, which commutes with the Hamiltonian, so that \eqref{eq:orginial_hamiltonina} is block-diagonal with respect to different symmetry sectors. We refer to $P$ as the ``parity" operator for the remainder of our discussion. When $g=0$, the ground state subspace can be thought of as the codespace of the quantum repetition code which we choose to be spanned by the symmetric and anti-symmetric Schr\"{o}dinger cat states given as,
\begin{align}\label{eq_ghz_def}
     \ket{\ghz_\pm}=\frac{\ket{0}^{\otimes N}\pm \ket{1}^{\otimes N}}{\sqrt{2}},
\end{align}
which are eigenvectors of parity with eigenvalues $\pm1$. The parity operator is the logical $\overline{X}$ operator of the quantum repetition code. Accordingly, the states in Eq. \eqref{eq_ghz_def} are the eigenstates of the logical  $\overline{X}$ in the code. 

Assume we start with a code state in the ground state subspace of the 1D Ising model at $g=0$,
\begin{align}\label{eq_original_code_state}
    \ket{\psi_0}=\alpha \ket{\ghz+}+\beta\ket{\ghz-}.
\end{align}
Then, the transverse field is increased slowly to a value $g<1$, over the course of time $T$. Let $\mathcal{U}_g$ denote the resulting adiabatic evolution operator. Since the Hamiltonian \eqref{eq:orginial_hamiltonina} is parity symmetric, $\mathcal{U}_g$ does not mix the even and odd symmetry sectors and hence one can follow evolution of each sector separately. Let $\ket{\varphi_+(g)}$ and $\ket{\varphi_-(g)}$ denote the ground states of $H_g$ in the positive and negative symmetry sectors, respectively. Based on the adiabatic theorem\cite{sakurai1995modern}, we have,
 \begin{align}\label{eq_adi_trasform_tfim}
     \mathcal{U}_g\ket{\ghz _\pm}=e^{i\eta_\pm}e^{i\xi_\pm}\ket{\varphi_\pm(g)},
 \end{align}
 where $\eta_\pm$ is the dynamical phase,
 \begin{align}
     \eta_\pm=-\frac{1}{\hbar}\int_0^T E_\pm(g(t)) \dd t
 \end{align}
 with $E_\pm(g)$ denoting the ground state energy in the $\pm$ symmetry sector of $H_g$ and $\xi_\pm$ is the geometric phase given as,
 \begin{align}
     \xi_\pm=i\int_0^T \mel{\varphi_\pm(g(t))}{\frac{\dd}{\dd t}}{\varphi_\pm(g(t))} \dd t.
 \end{align}
It is worth noting that since $T$ is finite, Eq. \eqref{eq_adi_trasform_tfim} holds only up to errors which are bounded by $1/T$, the inverse gap, $\norm{\frac{\dd}{\dd g} H}^2$ and $\norm{\frac{\dd^2}{\dd g^2} H}$ \cite{albash2018adiabatic}. Given that the gap remains finite and that $H_g$ is a sum of $O(N)$ local operators with bounded norms, assuming that $T$ grows parametrically faster than $N^2$  ensures that  Eq. \eqref{eq_adi_trasform_tfim} holds exactly in the thermodynamic limit. On the other hand, since in the ferromagnetic phase the ground state degeneracy is only lifted by exponentially small values in $N$, choosing $T$ to be at most $\text{poly}(N)$ ensures that $\eta_+=\eta_-=\eta$ up to a vanishing error. As for the geometric phase, we use the overall phase ambiguity in the definitions of $\ket{\varphi_\pm(g)}$ to enforce,
\begin{align}\label{eq_phase_fixing}
    \bra{\varphi_\pm(g)}\frac{\dd}{\dd g}\ket{\varphi_\pm(g)}=0,
\end{align} 
which then makes the geometric phases $\xi_\pm$ in Eq. \eqref{eq_adi_trasform_tfim} to vanish. Therefore, we find that under the adiabatic evolution, the original code state \eqref{eq_original_code_state} flows to,
\begin{align}\label{eq_adiabatic_image_of_psi0}
\ket{\psi_g}=\mathcal{U}_g\ket{\psi_0}=e^{i\eta}\Big[\alpha \ket{\varphi_+(g)}+\beta\ket{\varphi_-(g)}\Big],
\end{align}
The goal is to devise a recovery map $\mathcal{R}$ such that $\mathcal{R}(\ketbra{\psi_g})\simeq\ketbra{\psi_0}$, with an error that vanishes in the thermodynamic limit. As we shall argue in the following sections, the standard recovery map for the quantum repetition code fulfills this task within the entire ferromagnetic phase ($|g|<1$). To this end, first we review the standard decoder of the quantum repetition code.

\subsubsection{Standard Decoder for the Quantum Repetition Code}\label{sec_recovery_map}
The quantum repetition code can correct up to $N/2$ bit flip errors. 
The recovery map for the repetition code works as follows. First one measures the stabilizers $S_i=Z_i Z_{i+1}$ for $i=1,\cdots,N-1$. Let $s_i\in\{-1,1\}$ denote the outcome of the $S_i$ measurement. For convenience we define $m_i$ to denote the syndrome $m_i=(1-s_i)/2$ associated to $S_i$ measurement and $\bmm=(m_1,m_2,\cdots,m_{N-1})\in \mathbb{Z}_2^{N-1}$ to denote the collection of all syndromes. For any given $\bmm$, there are two possible Pauli-$X$ strings which are consistent with syndrome $\bmm$:
\begin{align}
    &e_{\bmm;1}=\prod_{i=1}^{N-1} X_i^{\sum_{j=i}^{N-1} m_i} \\
    &e_{\bmm;2}=P\,e_{\bmm;1}
\end{align}
Between $e_{\bmm;1}$ and $e_{\bmm;2}$, the decoder chooses the one with the lower weight\footnote{if $e_{\bmm;1}$ and $e_{\bmm;2}$ have the same weight, we pick $e_{\bmm;1}$}, which we denote by $e_{\bmm}$, and applies that operator to take the state back to the codespace. The recovery channel can thus be written as,
\begin{align}\label{eq_ising_recovery}
    \mathcal{R}(\rho)=\sum_{\bmm \in \mathbb{Z}_2^{N-1}}e_\bmm \Pi_\bmm \rho \Pi_\bmm e_\bmm,
\end{align}
where 
\begin{align}\label{eq_Pi_m_define}
    \Pi_\bmm=\prod_{i=1}^{N-1}\frac{1+(1-2m_i)Z_i Z_{i+1}}{2},
\end{align}
is the projection onto the subspace consistent with syndrome $\bmm$.
-

\subsubsection{Implementation of the Standard Decoder}
In this section we will first provide numerical evidence that the recovery map described in Section \ref{sec_recovery_map} can be used to recover the information throughout the ferromagnetic phase. In subsequent sections, we provide analytic arguments to explain the efficacy of the standard recovery map. For the numerical simulation, we take $\alpha=\beta=1/\sqrt{2}$ in Eq. \eqref{eq_original_code_state}, thus starting with,
\begin{align}
    \ket{\psi_0}=\frac{1}{\sqrt{2}}\ket{\ghz_+}+\frac{1}{\sqrt{2}}\ket{\ghz_-}.
\end{align}
The dynamical phase $e^{i\eta}$ in Eq. \eqref{eq_adiabatic_image_of_psi0} may be dropped as it is a global phase and has no bearing on the success of the decoder. Within each symmetry sector, $H_g$ can be written as a free fermion Hamiltonian using Jordan-Wigner transformation\cite{Jordan:1928wi}, which in turn allows us to find $\ket{\varphi_\pm(g)}$ analytically for any $g$. 
As is shown in Appendix \ref{apx_JWandFFM}, Eq. \eqref{eq_phase_fixing} can be satisfied by keeping the overlaps $\braket{\ghz_\pm}{\varphi_\pm(g)}$ real. For the sake of numerical simulation, we start directly with the state $\ket{\psi_g}=\frac{1}{\sqrt{2}}[\ket{\varphi_+(g)}+\ket{\varphi_-(g)}]$, instead of preparing $\ket{\psi_0}$ and simulating the adiabatic evolution. Next, we measure $Z_i Z_{i+1}$ stabilizers and choose the measurement outcomes based on the Born rule, decode the syndrome $\bmm$ and apply the inferred error string $e_\bmm$, to arrive at the error corrected state, $\ket{\psi'_0}=\frac{e_\bmm \Pi_\bmm\ket{\psi_g}}{\sqrt{\expval{\Pi_\bmm}{\psi_g}}}$. We estimate the fidelity of the recovery channel $\mathcal{F}=\mel{\psi_0}{\mathcal{R}(\ketbra{\psi_g})}{\psi_0}$, by averaging the overlap $|\braket{\psi'_0}{\psi_0}|^2$ over different runs with different measurement outcomes. We use the low-rank Gaussian state formalism\cite{bravyi2017complexity} to simulate the recovery process efficiently.

\begin{figure}
    \centering
    \includegraphics[width=\columnwidth]{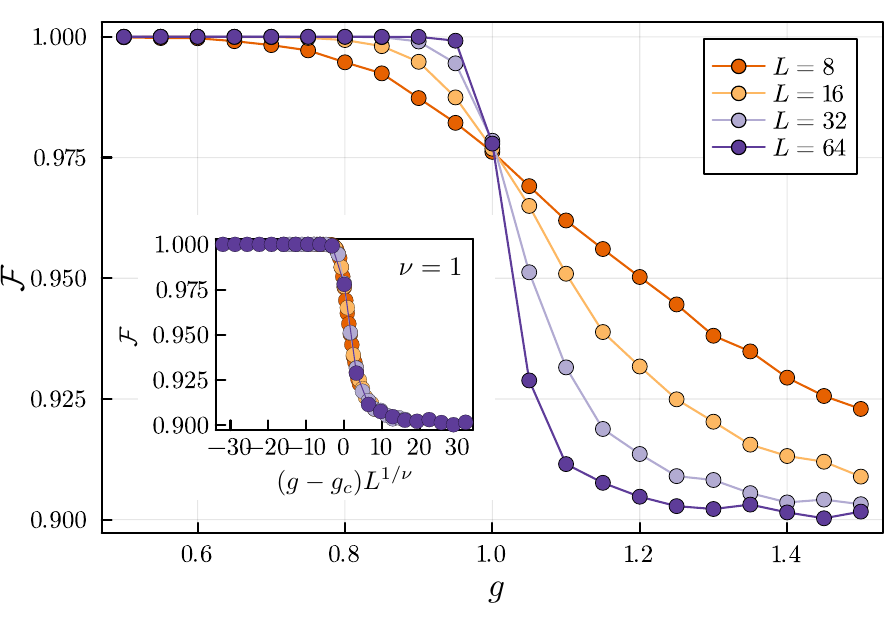}
    \caption{Fidelity of the recovered state with respect to the original encoded state as a function of the transverse field. The inset shows the data collapse of the same data points, using $g_c=1$ and $\nu=1$.}
    \label{fig_isingFidelity_c}
\end{figure}

Fig.~\ref{fig_isingFidelity_c} shows the average fidelity  $\mathcal{F}$ of the error-corrected state and the original encoded state, as a function of the transverse field $g$. As is clear from the figure, the fidelity approaches unity throughout the ferromagnetic phase in the thermodynamic limit. There is a decoding phase transition as one tunes the transverse field past the critical value into the paramagnetic phase. The inset shows the data collapse related to this phase transition, assuming the scaling form $\mathcal{F}(g,N)=F\Big((g-g_c)N^{1/\nu}\Big)$ with $\nu=1$ and $g_c=1$, which agree with the critical values corresponding to the order-disorder phase transition in the Ising model.

As mentioned earlier, we have ignored the energy difference of the symmetric and anti-symmetric ground states in the numerical simulations presented here. While this is justified in the ferromagnetic phase $g<1$, it is not a valid assumption for $g>1$. Therefore, one should ignore the $g>1$ part of the plot shown in Fig.~\ref{fig_isingFidelity_c}. In particular, while the fidelity in Fig.~\ref{fig_isingFidelity_c} seems to saturate to a large value $\sim 0.9$ in the paramagnetic phase, in realistic settings it would drop to zero past the critical point due to the random dynamical phases accumulated during the evolution which would decohere the information encoded in the relative phase of $\ket{\ghz_\pm}$.

\subsubsection{Syndrome Weights and the  Magnetization Order Parameter}\label{subsec_synd_and_order}
In this section we show that the distribution of syndromes when one measures the stabilizers $Z_i Z_{i+1}$ is related to the order parameter for the ferromagnetic phase. We will use this result,  alongside perturbative arguments in the following section, to explain why the standard decoder works perfectly in the ferromagnetic phase. 

The ferromagnetic phase is characterized by ferromagnetic long-range order (non-vanishing $Z_i Z_j$ correlations when $|i-j|\gg 1$). Accordingly, we can define the following operator, related to the square of the total magnetization, 
\begin{align}
    \hat{M}=\frac{1}{N(N-1)/2}\sum_{i<j}Z_i Z_j,
\end{align}
whose expectation value $M =\expval{\hat{M}}$ can be used as an order parameter for the ferromagnetic phase. Since $M$ is the average of $\expval{Z_i Z_j}$ over all possible $i\ne j$ pairs, its value approaches $\lim_{|i-j|\to\infty}\expval{Z_i Z_j}$ for large system sizes. 

\begin{figure}
    \centering
    \includegraphics[width=\columnwidth]{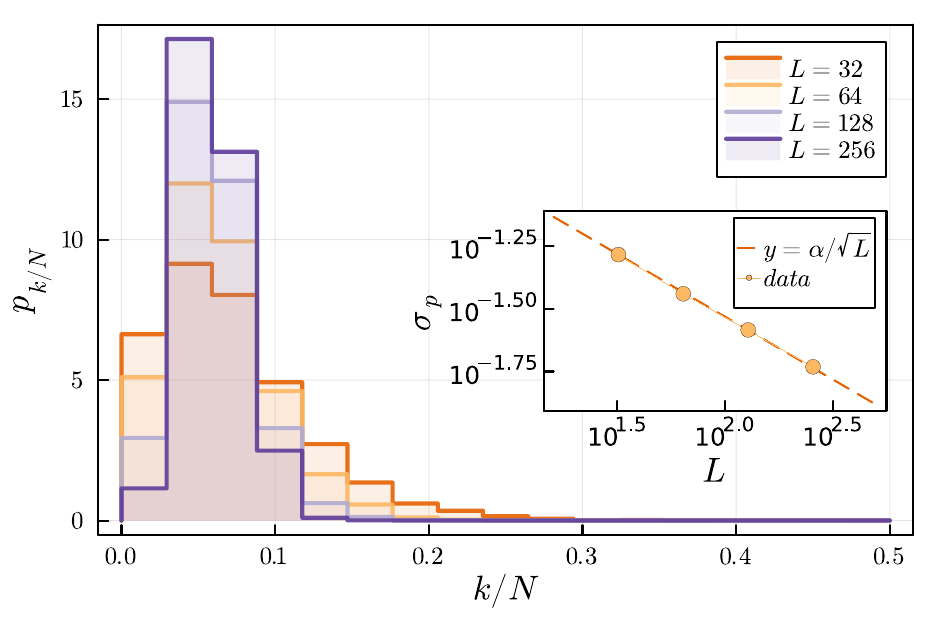}
    \caption{The probability distribution for the relative weight of the inferred error string $e_\bmm$ for $g=0.8$. As the system size grow, the distribution peaks around its mean. The inset shows the standard deviation of the distribution as a function of system size.}
    \label{fig_histo}
\end{figure}

Note that $\mathcal{B}=\cup_{\bmm \in \mathbb{Z}_2^{N-1}} \{e_\bmm\ket{\ghz_+},e_\bmm\ket{\ghz_-} \}$ is a complete set of independent eigenvectors of $\hat{M}$, with the corresponding eignevalues depending only on the size of non-trivial support of $e_\bmm$, which we denote by $k=|e_\bmm|$,
\begin{align}
    &\hat{M}~e_\bmm\ket{\ghz_\pm}=\frac{(N-2k)^2-N}{N^2-N}e_\bmm\ket{\ghz_\pm}.
\end{align}
Let us define $M_k$ to denote the eigenvalue,
\begin{align}\label{eq_mu_k}
    M_k=\frac{(N-2k)^2-N}{N^2-N}=\left(1-2\frac{k}{N}\right)^2+O(\frac{1}{N}),
\end{align}
so we may write $\hat{M}$ as,
\begin{align}\label{eq_mu_projection_expansion}
    \hat{M}=\sum_k M_{k} \Pi_k,
\end{align}
where $\Pi_k$ is the projection onto the subspace with eigenvalue $M_k$, which can written as
\begin{align}\label{eq_proj_k}
    \Pi_k=\sum_{\bmm:~|e_\bmm|=k} \Pi_\bmm,
\end{align}
with $\Pi_\bmm$ defined in Eq. \eqref{eq_Pi_m_define}. Thus, if one measures the order parameter $\hat{M}$ on the state $\ket{\psi_g}$, the probability that one finds the value $M_k$ is given by $\expval{\Pi_k}{\psi_g}$. On the other hand, when one measures all the stabilizers $Z_i Z_{i+1}$ for $i=1,\cdots,N-1$, the probability that the inferred error has weight $k$ is also given by $\expval{\Pi_k}{\psi_g}$. Therefore, the probability distribution for the weight of the inferred error $k$ is the same as the probability distribution for the measured value of the order parameter $M_k$.  Given that we expect the probability distribution for $M_k$ for a state in the ground state subspace of $H_g$ to be peaked around its mean value $M=\expval{\hat M}$ in the limit of large system size, Eq. \eqref{eq_mu_k}, along with the above reasoning, shows that the weight of the inferred error is related to the order parameter as,
\begin{align}\label{eq_k_typical}
    k=(1-\sqrt{M})\frac{N}{2},
\end{align}
with a relative error that goes to zero in the thermodynamic limit. This is easily checked numerically, as is shown in Fig.~\ref{fig_histo}.  
Interestingly, Eq. \eqref{eq_k_typical} shows that as one approaches the critical point where $M$ starts to vanish, the weight of the inferred error approaches $N/2$, which is exactly when one would expect the standard decoder to fail. However, due to the coherent nature of the errors, this simple picture is not complete and a more careful analysis is required to explain the perfect recovery in the ferromagnetic phase. In particular, after measuring the stabilizers and say finding syndrome $\bmm$, the state of the system would be
\begin{align}\label{eq_pm_psi}
    \Pi_\bmm \ket{\psi_g}=&\alpha~e_\bmm \ket{\ghz_+}\mel{\ghz_+}{e_\bmm}{\varphi_+(g)}\nonumber\\
    &+\beta~e_\bmm \ket{\ghz_-}\mel{\ghz_-}{e_\bmm}{\varphi_-(g)},
\end{align}
up to a normalization factor. As we show in the next section, the local indistinguishably of the ground states, ensures that the coherent factors $\mel{\ghz_\pm}{e_\bmm}{\varphi_\pm(g)}$ in Eq. \eqref{eq_pm_psi} are equal up to exponentially small corrections, so syndrome measurements practically decohere the errors, and finite value of the order parameter $M$ then ensures the success of the recovery map. 
\subsubsection{Perturbative Argument for the Success of the Decoder}
In this section, we will use perturbation theory to explain the success of the standard recovery map in the ferromagnetic phase. Let $\ket{\psi_0}$ and $\ket{\psi_g}$ denote the original code state and its image under adiabatic evolution, respectively (Eq. \eqref{eq_original_code_state} and Eq. \eqref{eq_adiabatic_image_of_psi0}). 
The fidelity of the corrected state and the encoded state can be expressed as,
\begin{align}
    &\mel{\psi_0}{\mathcal{R}\Big(\ketbra{\psi_g}\Big)}{\psi_0}=\sum_\bmm |\mel{\psi_0}{e_\bmm \Pi_\bmm}{\psi_g}|^2\nonumber\\
    &=\sum_\bmm |\mel{\psi_0}{e_\bmm}{\psi_g}|^2
    =\sum_\bmm \Bigg|\,|\alpha|^2\,\mel{\ghz_+}{e_\bmm }{\varphi_+(g)}\nonumber \\
    &+|\beta|^2\,\mel{\ghz_-}{e_\bmm}{\varphi_-(g)}\Bigg|^2,\label{eq_fidelity_sum}
\end{align}
where we have used the following,
\begin{align}
    e_\bmm \Pi_\bmm=\Pi_{\mathbf{0}}e_\bmm,
\end{align}
and the fact that $\Pi_{\mathbf{0}}\ket{\psi_0}=\ket{\psi_0}$. As is shown in Appendix \ref{apx_perturb_theory_TFIM}, by explicitly writing down the perturbative expansions of $\ket{\varphi_\pm(g)}$, one can show that
\begin{align}\label{eq_pm_equal_mel}
   &\frac{\mel{\ghz_-}{e_\bmm}{\varphi_-(g)}}{\mel{\ghz_+}{e_\bmm}{\varphi_+(g)}} =1+\mathcal{O}(g^{(\frac{1}{2}-\frac{k}{N})N}),
\end{align}
where $k=|\text{supp}(e_\bmm)|\le N/2$. Eq. \eqref{eq_pm_equal_mel} follows from the observation that while $e_\bmm\ket{\ghz_+}$ appears in the perturbative expansion at the order $k$ of perturbation theory, it is only the terms of order $N/2$ and beyond that see the difference between the odd and even parity sectors. If we plug this expression into Eq. \eqref{eq_pm_equal_mel} we find that,
\begin{align}\label{eq_f_lowerbound}
    &\mel{\psi_0}{\mathcal{R}\Big(\ketbra{\psi_g}\Big)}{\psi_0}\ge 1-\varepsilon,
\end{align}
with 
\begin{align}
    \varepsilon&=\sum_\bmm |\mel{\ghz_+}{e_\bmm }{\varphi_+(g)}|^2 \mathcal{O}(g^{(\frac{1}{2}-\frac{k}{N})N})\nonumber.
\end{align}
To arrive at Eq. \eqref{eq_f_lowerbound} we used $\sum_\bmm |\mel{\ghz_+}{e_\bmm}{\varphi_+(g)}|^2=1$, which follows from the fact that $ \{ e_\bmm\ket{\ghz_+}\,|\,\bmm \in \mathbb{Z}_2^{N-1}\}$ is a complete basis for the positive parity sector. Let us define $p_k$ as,
\begin{align}
    p_k&=\sum_{\bmm:~|e_\bmm|=k} |\mel{\ghz_+}{e_\bmm }{\varphi_+(g)}|^2\nonumber\\
    &=\mel{\varphi_+(g)}{\Pi_k}{\varphi_+(g)},
\end{align}
where $\Pi_k$ is defined in Eq. \eqref{eq_proj_k}. Accordingly, one may write the error $\varepsilon$ as,
\begin{align}\label{eq_err_full}
    \varepsilon=\sum_{k=1}^{N/2} p_k~\mathcal{O}(g^{(\frac{1}{2}-\frac{k}{N})N}).
\end{align}
As we argued in the previous section, $p_k$ will be peaked around $\frac{k}{N}=\frac{1}{2}-\frac{\sqrt{M}}{2}$, where $M=\expval{\hat{M}}{\varphi_+(g)}>0$ is the order parameter of the ferromagnetic phase. Thus we can drop the tail of the sum in Eq. \eqref{eq_err_full} by the virtue of $p_k$ going to $0$ for $k$ far from its mean value. By doing so, we find $\varepsilon=\mathcal{O}(\exp(-\sqrt{M}N))$, and thus,
\begin{align}
    &\mel{\psi_0}{\mathcal{R}\Big(\ketbra{\psi_g}\Big)}{\psi_0}\ge 1-\mathcal{O}(\exp(-\sqrt{M}N)),
\end{align}
with $M>0$ in the ferromagnetic phase $|g|<1$.

It is worth noting that neither the arguments in Section \ref{subsec_synd_and_order} nor the perturbative arguments in this section depend on the integrability of the TFIC. Hence we expect that our result would remain true in more generic systems as well. As an example, in Appendix \ref{apx_nonintg_tfim} we show that the decoder works perfectly up to a phase transition, even when the integrability is broken. 

Moreover, while the analysis in this section has been focused on a spontaneous symmetry broken phase, the same observation can be made with regard to a wide class of one-dimensional symmetry protected topological phases. As an example, in Appendix \ref{apx_clustermodel}, we study the  $\mathbb{Z}_2\times \mathbb{Z}_2$ cluster state Hamiltonian on a open chain and in presence of a transverse field, and we show that the standard recovery map succeeds with probability $1$ throughout the topological phase. In this case, the SPT string order parameter $\expval{s_{ij}}$ plays the same role as the Ising order parameter $\expval{Z_i Z_j}$. We expect the same result to hold more generally for any zero form SPT phase protected by a  finite Abelian symmetry group.


\subsection{Perturbed Toric Code}\label{sec_toric_code}
Next,  we consider the adiabatic drift inside the phase diagram of the toric code under a special perturbation which allows for the perturbed ground states to be obtained exactly throughout the phase diagram \cite{castelnovo2008quantum}. Consider the $\Lambda=L\times L$ square lattice on a cylinder, with rough boundary on top and bottom edges and with qubits placed on the edges. The toric code Hamiltonian is defined as,
\begin{align}
    H_\text{TC}=-\sum_p B_p -\sum_s A_s,
\end{align}
where $p$ and $s$ correspond to plaquettes  and stars of $\Lambda$ and $B_p=\prod_{i\in p}Z_i$ and $A_s=\prod_{i\in s}X_i$ are the  plaquette and star operators of the toric code. The plaquette operators on the top and bottom boundary include three qubits, while the rest of plaquette operatros as well as all of the star operators act on four qubits. $H_\text{TC}$ has doubly degenerate ground states, which we denote by $\ket{\varphi_i(0)}$. They are eigenstates of the logical $\overline{Z}$ operator, 
\begin{align}
   \overline{Z} \ket{\varphi_i(0)}=(-1)^i\ket{\varphi_i(0)}, \quad i=0,1,
\end{align}
where $\overline{Z}$ is the Pauli-$Z$ string that runs from top to bottom boundary. Consider the following perturbed toric code Hamiltonian
\begin{align}\label{eq_CC4_hamiltonian}
    H(\beta)=H_\text{TC} +\sum_s \Bigg[\exp(-\beta \sum_{j\in s}Z_j)-1\Bigg],
\end{align}
where the sum on $s$ runs over the stars of the lattice $\Lambda$. Notably, when $\beta \ll 1$ the Hamiltonian \eqref{eq_CC4_hamiltonian} reduces to the Hamiltonian of the toric code under uniform magnetic field in $Z$ direction, with the strength of the magnetic field proportional to $\beta$.  It is shown in Ref. \cite{castelnovo2008quantum} that the topological phase extends up to $\beta_c=\frac{1}{2}\ln(\sqrt{2}+1)\simeq 0.44$, at which point the system undergoes a second order phase transition within the two-dimensional Ising universality class, and beyond which the topological order is lost. This phase transition is not generic; this model provides an example of a conformal quantum critical point \cite{ardonne2004topological,henley2004classical}, where the ground-state wave function itself exhibits two-dimensional conformal symmetry when $\beta = \beta_{c}$, while a time-independent conformal symmetry is absent at a generic quantum critical point. Nevertheless, the nature of the gapped phase far from the critical point, is sufficiently generic.  For example, the gapped phase exhibits a quantized topological entanglement entropy \cite{levin2006detecting,kitaev2006topological}, which remains at $2\log(2)$ for $\beta<\beta_c$, as is characteristic of the toric code phase, and vanishes when $\beta>\beta_c$. 

\begin{figure}
 \begin{subfigure}{0.49\columnwidth}
     \includegraphics[width=0.7\textwidth]{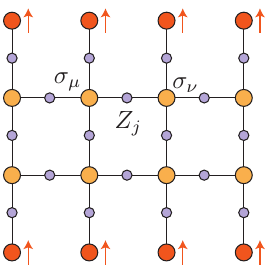}
     \caption{}
     \label{fig:2dising_toric}
 \end{subfigure}
 \hfill
 \begin{subfigure}{0.49\columnwidth}
     \includegraphics[width=0.75\textwidth]{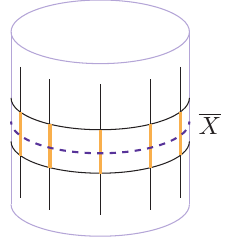}
     \caption{}
     \label{fig:ising_domainwall}
 \end{subfigure}
 \caption{(a) The qubits (small circles) are on the edges of $\Lambda$ while the classical Ising spins (large circles) sit on the vertices. The spins on the top and bottom boundary (dark large circles) are fixed to the $\sigma=+1$ state (b) $\calZ_0(\beta)$ equals the partition function of 2D-classical Ising model on $\Lambda$ with fixed boundary conditions on top and bottom. $\calZ_1(\beta)$ is equal to the same partition function but with a domain wall inserted along the support of $\overline{X}$, which changes the coupling on highlighted links to be anti-ferromagnetic. }
 \label{Label}
\end{figure}

The ground states of $H(\beta)$ remain two-fold degenerate, and can be expressed exactly as \cite{castelnovo2008quantum},
\begin{align}\label{eq_C4_gs}
    \ket{\varphi_i(\beta)}=\frac{1}{\calZ_i(\beta)^{\frac{1}{2}}}\exp(\frac{\beta}{2}\sum_j Z_j)\ket{\varphi_i(0)}, \quad i=0,1, 
\end{align}
where $\calZ_i(\beta)=\expval{\exp(\beta\sum_j Z_j)}{\varphi_i(0)}$ is the normalization factor which can be expressed as the partition function of 2D classical Ising model. In particular, $\calZ_0(\beta)$ is given as (see Appendix \ref{apx_calZ_derivation} for its derivation)
\begin{align}\label{eq_C4_partition_0}
    \calZ_0(\beta)=\frac{1}{2^M}\sum_{\sigma\in\{-1,1\}^M}\exp(\beta \sum_{\langle \mu,\nu \rangle} \sigma_\mu \sigma_\nu),
\end{align}
where $\sigma_\mu=\pm 1$ are classical spins which are placed on the stars of the lattice $\Lambda$. We also put fixed $\sigma_\mu=+1$ spins at the top and bottom open boundaries of $\Lambda$ (see Fig.\ref{fig:2dising_toric}). $\calZ_1(\beta)$ is  the same partition function with a domain wall along the support of  $\overline{X}$ (see Fig.~\ref{fig:ising_domainwall}), along which the Ising coupling is flipped to be anti-ferromagnetic.

\subsubsection{Adiabatic Evolution}
Imagine we start with the $\ket{\varphi_i(0)}$ state and tune $\beta$ from $0$ to a finite value $\beta<\beta_c$. First, note that since $\overline{Z}$  commute with the Hamiltonian $H(\beta)$, its value will not change under the adiabatic evolution. This shows that $\ket{\varphi_i(0)}$ will be mapped to $\ket{\varphi_i(\beta)}$ up to dynamical and geometric phases associated with the adiabatic evolution, analogous to Eq. \eqref{eq_adi_trasform_tfim} in the TFIC example. Since the ground states remain exactly degenerate, the dynamical phase is a constant independent of $i$ and hence we may simply ignore it. As for the geometric phase, it is given by the integral of $\mel{\varphi_i(\beta')}{\frac{\dd}{\dd \beta'}}{\varphi_i(\beta')}$, which can be computed from expression \eqref{eq_C4_gs},
 \begin{align}
     &\mel{\varphi_i(\beta)}{\frac{\dd}{\dd \beta}}{\varphi_i(\beta)}\nonumber\\
     =&\frac{-1}{2}\frac{\partial_\beta \calZ_i}{\calZ_i^2}\expval{\exp(\beta \sum_j Z_j)}{\varphi_i(0)}\nonumber\\
     &+ \frac{1}{\calZ_i} \expval{\frac{1}{2}\sum_k Z_k\exp(\beta \sum_j Z_j)}{\varphi_i(0)}\nonumber\\
     =&\frac{-1}{2}\frac{\partial_\beta \calZ_i}{\calZ_i^2}\calZ_i+\frac{1}{2} \frac{1}{\calZ_i} \partial_\beta \calZ_i = 0,
\end{align}
This discussion shows that $\ket{\varphi_i(\beta)}$ as defined in Eq. \eqref{eq_C4_gs} can be taken to be the image of $\ket{\varphi_i(0)}$ under adiabaticly tuning $\beta$ from $0$ to $\beta$. 

\subsubsection{Recovery with Stabilizer Measurements and Pauli Feedback}
Consider encoding a state in the codespace of $H_\text{TC}$ and then tuning $\beta$ from $0$ to a finite value $\beta$. Assuming the value of $\beta$ is known, we show that by measuring stabilizers and  applying an appropriate Pauli feedback one can recover the original state with perfect fidelity in the thermodynamic limit, as long as $\beta<\beta_c$.  

Given a syndrome data, there will be two different classes of Pauli strings which would correct the syndrome and move the state back to the codespace. Pauli strings that belong to the same class, map the corrupted state to the same state in the code state, while the action of Pauli strings from different classes, differ by the application of a logical operator. The task of a Pauli decoder is to choose a class to apply a Pauli string from that class to correct the syndromes. The minimum weight perfect matching (MWPM) decoder chooses the class that contains the Pauli string with the minimum weight. For incoherent Pauli noise models, the best performance is achieved by the maximum likelihood (ML) decoder which chooses the class which is the most probable to contain the Pauli error that actually occurred\cite{dennis2002topological}. Another decoder which is optimal in the thermodynamic limit as well For coherent noise consider here, one may define the maximum likelihood decoder to be the decoder that chooses the class which maximizes the expected fidelity between the recovered state and the original encoded state, assuming a uniform distribution in the codespace for the encoded state. Here we consider a probabilistic likelihood (PL) decoder, which assigns certain weights to each class and then chooses a class randomly using the assigned weights. If the noise was incoherent and if the weights were chosen to be equal to the likelihood of each class, PL decoder would perform as good as the ML decoder in the thermodynamic limit.

Let $\ket{\psi_0}$ denote the encoded state, 
\begin{align}
    \ket{\psi_0}=\alpha_0 \ket{\varphi_0(0)}+\alpha_1 \ket{\varphi_1(0)},
\end{align}
with $|\alpha_0|^2+|\alpha_1|^2=1$. Under adiabatic quantum evolution, this state gets mapped to 
\begin{align}\label{eq_toric_psi_beta}
    \ket{\psi_\beta}=\alpha_0 \ket{\varphi_0(\beta)}+\alpha_1 \ket{\varphi_1(\beta)},
\end{align}
up to a global phase which we ignore. We are interested in recovering $\ket{\psi_0}$ by measuring stabilizers on $\ket{\psi(\beta)}$ and applying a Pauli feedback.

The adiabatically evolved basis states can be written as,
\begin{align}
    \ket{\varphi_i(\beta)}&=\frac{1}{\calZ_i(\beta)^{\frac{1}{2}}}\exp(\frac{\beta}{2}\sum_j Z_j)\ket{\varphi_i(0)}\\
    =&\frac{\cosh(\frac{\beta}{2})^{N}}{\calZ_i(\beta)^{\frac{1}{2}}}\sum_{e\in\mathbb{Z}_2^N}\tanh(\frac{\beta}{2})^{|e|}Z(e) \ket{\varphi_i(0)}\label{eq_psibeta_1},
\end{align}
where $e$ is a $N$-element binary vector, $|e|$ is its Hamming weight and $Z(e)$ is the Pauli-$Z$ error string corresponding to the $e$ binary vector, i.e.
\begin{align}
    Z(e)=\prod_{j}Z_j^{e_j}.
\end{align}
Define the parameter $q$ as,
\begin{align}\label{eq_q_vs_beta}
    q=\frac{\sinh(\beta/2)}{\sinh(\beta/2)+\cosh(\beta/2)}=\frac{1-e^{-\beta}}{2},
\end{align}
so we may write Eq. \eqref{eq_psibeta_1} as,
\begin{align}
    \ket{\varphi_i(\beta)}=\frac{1}{\calN_i^{\frac{1}{2}}}\sum_{e\in \mathbb{Z}_2^N}(1-q)^{N-|e|}q^{|e|}Z(e)\ket{\varphi_{i}(0)}\label{eq_psibeta_2},
\end{align}
$P([e])$ as,
\begin{align}
    P([e])=\sum_{e'\in [e]}(1-q)^{N-|e'|}q^{|e'|}.
\end{align}
In a hypothetical incoherent noise model where each qubit undergoes a phase flip with probability $q$, $P([e])$ denotes the probability that the resulting error is in the equivalency class $[e]$. However since in our setup the noise is coherent, we are using $P([e])$ only as an ancillary function rather than representing the probability of a specific event in our setup. For a given syndrome $\bmm$, there are many errors $e$ that are compatible with $\bmm$, meaning that measuring star operators on $Z(e)\ket{\psi_0}$ gives syndrome $\bmm$. Let $e_\bmm$ denote the error among them for which $P([e_\bmm])$ is maximum. Since $P([e])$ is only a function of the class $[e]$, there are many equivalent choices of $e_\bmm$ but for our discussion, it does not matter which one is chosen as $e_\bmm$\footnote{ In fact we may choose $e_\bmm$ to be any string that is compatible with $\bmm$ and fix that choice throughout. Using the output of ML decoder is just an easy way to fix that choice.}. In the hypothetical incoherent phase flip noise model with rate $q$, $e_\bmm$ is the error string that would be chosen by ML decoder given a syndrome $\bmm$. Using this notation we may rewrite Eq. \eqref{eq_psibeta_2} as, 
\begin{align}\label{eq_psibeta_3}
    \ket{\varphi_i(\beta)}=\frac{1}{\sqrt{\calN_i}}\sum_{\bmm,k}(-1)^{k\,i}P([e_\bmm+\ell_k])Z(e_\bmm)\ket{\varphi_i(0)},
\end{align}
where, above and hereafter in this section, $\bmm$ and $k$ sum over $\mathbb{Z}_2^M$ and $\{0,1\}$ respectively.

The PL decoder is defined as follows: Given the state $\ket{\psi_\beta}$ in Eq. \eqref{eq_toric_psi_beta}, first we measure all the stabilizers, obtaining an specific outcome, say $\bmm$, then we choose $k=0,1$ randomly according to probabilities $\frac{P([e_{\bmm}+\ell_k])}{\sum_{k'} P([e_{\bmm}+\ell_{k'}])}$ and apply $Z(e_\bmm + \ell_k)$. The fidelity of the resulting state and the original encoded state $\ket{\psi_0}$, averaged over possible syndrome outcomes has the following form,
\begin{align}\label{eq_fidelity_sum_toric_example}
    \mathcal{F}=&\sum_{\bmm,k}\frac{P([e_\bmm+l_k])}{\sum_{k'}P([e_\bmm+l_{k'}])}~\Bigg[\nonumber\\
    &\frac{|\alpha_0|^4}{\calN_0}(P([e_\bmm])+P([e_\bmm+\ell_1]))^2\nonumber\\
    &+\frac{|\alpha_1|^4}{\calN_1}(P([e_\bmm])-P([e_\bmm+\ell_1]))^2\nonumber\\
    &+(-1)^k 2\frac{|\alpha_0|^2 |\alpha_1|^2}{\calN_0^{\frac{1}{2}} \calN_1^{\frac{1}{2}}}(P([e_\bmm])^2-P([e_\bmm+\ell_1])^2)\Bigg].
\end{align}
First, consider the case where $\alpha_0=1$ and $\alpha_1=0$. Note that in this case $\mathcal{F}=1$ trivially. This is because $\ket{\psi_\beta}=\ket{\varphi_0(\beta)}$ is the eigenvector of $\overline{Z}$ and stabilizer measurements and the applied $Z$ feedback will not change that, so the recovered state has to be $\ket{\varphi_0(0)}$. This shows that, 
\begin{align}\label{eq_N_0_sum}
    \sum_{\bmm}(P([e_\bmm])+P([e_\bmm+\ell_1]))^2=\calN_0
\end{align}
We remark that one could explicitly calculate the sum on the left hand side, by mapping the sum to the partition function of an associated classical 2D random-bond Ising model\cite{dennis2002topological,chubb2021statistical} which of course ends up to be equal to $\calN_0$. The details of this calculation is included in Appendix \ref{apx_explicit_summation} for completeness. 
Based on a similar reasoning when starting with $\alpha_0=0$ and $\alpha_1=1$, we find, 
\begin{align}\label{eq_N_1_sum}
    \sum_{m}(P([e_\bmm])-P([e_\bmm+\ell_1]))^2=\calN_1.
\end{align}
We need to calculate the third sum in Eq. \eqref{eq_fidelity_sum_toric_example} to find the average fidelity when both $\alpha_0$ and $\alpha_1$ are non-zero. However the third sum is the same as Eq. \eqref{eq_N_1_sum},
\begin{align}
    &\sum_{\bmm,k}\frac{P([e_\bmm+l_k])(-1)^k}{\sum_{k'}P([e_\bmm+l_{k'}])}(P([e_\bmm])^2-P([e_\bmm+\ell_1])^2)\nonumber\\
    &=\sum_{\bmm}\frac{(P([e_\bmm])-P([e_\bmm+\ell_1]))^2}{\sum_{k'}P([e_\bmm+l_{k'}])}(P([e_\bmm])+P([e_\bmm+\ell_1]))\nonumber\\
    &=\sum_{\bmm} (P([e_\bmm])-P([e_\bmm+\ell_1]))^2=\calN_1.
\end{align}
Therefore, we find that in general, the average decoding fidelity would be,
\begin{align}
    \mathcal{F}=|\alpha_0|^4+|\alpha_1|^4+2|\alpha_0|^2 |\alpha_1|^2 \sqrt{\frac{\calZ_1(\beta)}{\calZ_0(\beta)}}.
\end{align}
When $\beta<\beta_c$ the 2D Ising model is in the disordered phase and hence $\calZ_1=\calZ_2$, up to errors that would vanish in the thermodynamic limit. Therefore, $\mathcal{F}$ approaches $1$ in the thermodynamic limit.

\section{Reversing an Adiabatic Noise Channel}\label{sec_adiabatic_channel}
We now consider reversing an adiabatic noise channel as defined in Section \ref{sec_general_setup}. The main difference between this setup and the setup considered in Section \ref{sec_rev_adi_evo} is that here we assume no knowledge of the adiabatic path $\gamma$ along which the system has been evolved. In particular, we assume the encoded state $\ket{\psi_0}$ is acted upon by the following quantum adiabatic channel, 
\begin{align}
    \widetilde{\rho}=\mathcal{E}_\Gamma (\ketbra{\psi_0})=\sum_{\gamma\in\Gamma}p_\gamma \mathcal{U_\gamma} \ketbra{\psi_0} \mathcal{U_\gamma}^\dagger,
\end{align}
where $\Gamma$ is a set of paths $\gamma(t)\in \mathcal{A}$ with $\mathcal{A}$ a neighborhood of of the code Hamiltonian inside the gapped phase, and $p_\gamma$ is a probability over $\Gamma$. 

First we briefly revisit the examples considered in Section \ref{sec_rev_adi_evo} and comment on the recovery process when the adiabatic path is not known. Then we provide a simple example to show that in general it might be impossible to recover the information under $\mathcal{E}_\Gamma$, even if $\mathcal{A}$ is far from the gapped phase boundaries. Lastly, we derive sufficient conditions for the existence of a recovery channel after the application of the adiabatic noise $\mathcal{E}_\Gamma$, by studying the ``decoupling" of the environment from reference qubits which are maximally-entangled with the codespace.  

\subsection{Recovery for the TFIC and the Perturbed Toric Code}\label{sec_ising_and_toric_as_channels}
As was noted earlier, the recovery map for TFIC described in Section \ref{sec_repetition_code_and_1dTFIM} does not use any information about the adiabatic path $\gamma$. This readily shows that the same recovery map can perfectly reverse an adiabatic noise channel $\mathcal{E}_\Gamma$ when acting on the codespace, as long as $\mathcal{A}$ is contained entirely within the ferromagnetic phase of the TFIC.

On the other hand, the PL decoder used to reverse an adiabatic evolution in the perturbed toric code example in section \ref{sec_toric_code} explicitly makes use of the value of $\beta=\gamma(T)$. Therefore the PL decoder is no longer readily applicable to the case where $\gamma$ is unknown. A possible solution is to use the MWPM decoder, which does not depend on $\beta$. To analyze the performance of MWPM decoder, it helps to have a better understanding of the effective noise model after stabilizer measurements. 

Let us focus for now on a single trajectory $\gamma$ instead of an ensemble $\Gamma$ of trajectories. The fact that PL decoder is in principle able to recover the original state with perfect fidelity after stabilizer measurements already proves that the effective noise channel describing the combined effect of an adiabatic evolution followed by stabilizer measurements is a Pauli noise model. In other words, stabilizer measurements effectively decohere the coherent noise associated to $\mathcal{U}_\gamma$ evolution. However, the resulting Pauli noise channel is not necessarily local, i.e. it may not be expressible by a tensor product of local Pauli noise channels. A useful probe for the locality of a noise model is the correlations between the syndromes. Let $m_s,m_{s'}\in\{-1,1\}$ denote the random variables corresponding to the measurement outcomes of stabilizers $A_s$ and $A_{s'}$. The connected correlation between the syndromes at $s$ and $s'$ is 
\begin{align}
    C(s,s')=\overline{m_s m_{s'}}-\overline{m_s}\times\overline{m_{s'}},
\end{align}
where the line denotes an average over the random measurement outcomes. For strictly local incoherent noise models such as i.i.d. Pauli noise, as well as correlated Pauli noise models with locally factored distributions \cite{chubb2021statistical}, the syndromes are completely uncorrelated beyond a certain finite length, so that $C(s,s')=0$ when the distance between $s$ and $s'$ is larger than this length-scale. For a generic, locally-correlated noise model, one expects $C(s,s')$ to decay as a function of $|s-s'|$, with a decay form and characteristic length scale that is set by the correlations in the noise model. As is shown in Appendix \ref{apx_synd_corr}, in the ground state subspace of $H(\beta)$, $C(s,s')$ is given as,
\begin{align}\label{eq_synd_corr}
    C(s,s')=\overline{\exp(\beta \epsilon_s)\exp(\beta\epsilon_{s'})}-\overline{\exp(\beta \epsilon_s)}\times \overline{\exp(\beta \epsilon_{s'})},
\end{align}
where the average has been taken with respect to the 2D classical Ising model at temperature $\beta$  with the classical spins $\sigma_\mu=\pm 1$ living on the stars, and $\epsilon_s=-\sigma_s\sum_{\langle \mu, s\rangle}\sigma_\mu$ is the local energy at site $s$. Based on physical grounds we expect $C(s,s')$ to have the following form,
\begin{align}\label{eq_synd_corr_exponential}
    C(s,s')\sim \exp(-|s-s'|/\xi(\beta)),
\end{align}
where $\xi(\beta)$ is the correlation length in the Ising model. Therefore deep inside the topological phase corresponding to $\beta\ll \beta_c$, the correlation length is small (of the order of the lattice spacing) and the effective noise model is a weak, local Pauli channel, and as such we expect the MWPM decoder to succeed. However,  for larger values of $\beta$, not only does the noise strength increase, but the noise correlations become non-local, as the correlation length increases and eventually diverges at the critical point $\beta=\beta_c$. Therefore we expect the MWPM decoder -- which has no prior knowledge of the noise correlations -- to generically have a finite threshold $\beta_\text{th}^\text{MWPM}$ strictly less than $\beta_c$, where the PL decoder fails, which means that one can reverse $\mathcal{E}_\Gamma$ using MWPM only if $\mathcal{A}$ is contained in $[0,\beta_\text{th}^\text{MWPM}]$, which is strictly smaller than the topological gapped phase.

\subsection{Rotated Ising Model}\label{sec_ising_rotation}
In this section we provide a simple example to show that when the taken adiabatic path is not known, remaining in the same quantum phase does not guarantee the ability to recover the original information encoded in the ground state subspace. 
 
 Let $ H_0=-\sum_{i=1}^N Z_{i}Z_{i+1}$ denote the the 1D Ising Hamiltonian on a closed ring of $N$ qubits, whose ground state subspace can be thought of as the codespace of the quantum repetition code, spanned by $\ket{\ghz_\pm}$, as was explained in Section \ref{sec_repetition_code_and_1dTFIM}. 
 Consider the following family of Hamiltonians,
 \begin{align}\label{eq_counter_exp}
     H_\theta=U_\theta H_0 U_\theta^\dagger, \qquad U_\theta=\prod_{j=1}^N e^{-i\,\frac{\theta}{2}\,X_j},
 \end{align}
 for $-\pi< \theta\le \pi$. Since $U_\theta$ is a tensor product of local terms, $H_\theta$ remains a local Hamiltonian for any $\theta$. Moreover, because $U_\theta$ is unitary,  $H_\theta$ has the same energy spectrum as $H_0$ and thus the energy gap remains open as one deforms $H_0$ to $H_\theta$. The ground states of $H_\theta$ are 
 \begin{align}\label{eq_rotated_ising_gs}
     \ket{\varphi_\pm(\theta)}=U_\theta \ket{\ghz_\pm}.
 \end{align}
 Consider starting from $\ket{\ghz_+}$ and adiabatically tunning $\theta$ form $0$ to a finite value $\theta$. Since $U_\theta$ is symmetric, the adiabatic evolution does not mix the even and odd symmetry sectors of the Hamiltonian, and hence one can follow evolution of each sector separately. This shows that the image of $\ket{\ghz_+}$ under adiabatic evolution would be $\ket{\varphi_+(\theta)}$ as is given by Eq. \eqref{eq_rotated_ising_gs}, up to a dynamical phase and a geometric phase, similar to the expression \eqref{eq_adi_trasform_tfim} for the TFIC. Analogously, $\ket{\ghz_-}$ will evolve into  $\ket{\varphi_-(\theta)}$ up to a phase. Since the ground states remain exactly degenerate as one tunes $\theta$, the accumulated dynamical phase is exactly equal for the two ground states and thus can be neglected as global phase. As for the geometric phase, note that 
  \begin{align}
     \mel{\varphi_\pm(\theta)}{\frac{\dd}{\dd\theta}}{\varphi_\pm(\theta)}=-\frac{i}{2}\sum_{i=1}^N\mel{\ghz_\pm}{X_i}{\ghz_\pm}=0.
 \end{align}
Therefore, $\ket{\varphi_\pm(\theta)}$ states as given by Eq. \eqref{eq_rotated_ising_gs} can be taken as the image of $\ket{\ghz_\pm}$ under adiabatic evolution.

Let $P=\Pi_j X_j$ denote the parity operator, which is a logical operator of the repetition code. Note that
\begin{align}
    U_{\pi}=(-i)^{N}\prod_{i=1}^N X_i\propto P,
\end{align}
Therefore, tuning $\theta$ from $0$ to $\pi$ adiabatically, while keeps the system in the same phase, maps a code state $\ket{\psi_0}=\alpha \ket{\ghz_+}+\beta\ket{\ghz_-}$ to another code state $\ket{\psi_\pi}=\alpha \ket{\ghz_+}-\beta\ket{\ghz_-}$, up to a global phase, and hence implements a logical operation on the codespace which no decoder can detect and correct. More generally, note that
\begin{align}
    U_{\pi-\theta}=(-i)^N P\times U_{-\theta}.
\end{align}
This shows that if there are paths $\gamma_{-\theta}$ and $\gamma_{\pi-\theta}$ in $\Gamma$ which end at $-\theta$ and $\pi-\theta$ respectively, no decoder can perfectly reverse $\mathcal{E}_\Gamma(\ketbra{\psi_0})$. Nonetheless, perfect recovery might be possible if one restricts $\theta$ to a small neighborhood of $0$ such that either $-\theta$ or $\pi-\theta$ is in that neighborhood but not both. Considering only the symmetric intervals around $0$, the largest neighborhood $\mathcal{A}$ for which perfect recovery could be possible is $-\pi/2<\theta<\pi/2$. Interestingly, perfect recovery is in fact possible in this case, as is shown in Appendix \ref{apx_decoding_rotated_ising_model}.

While this example is concerned with a symmetry breaking phase of matter, it is clear that one can easily build similar examples in concerning topological error correcting codes which are fixed points of topological phases of matter. After all, topological quantum computation is based on the idea of implementing logical gates on the codespace via adiabatically deforming the Hamiltonian along a non-trivial cycles\cite{kitaev2003fault,nayak2008non}.

\subsection{General Conditions for the Existence of a Recovery Channel}\label{sec_coherent_info}
In this section we address the existence of a recovery map for the adiabatic noise channel in a general gapped phase as described in Section \ref{sec_general_setup}, instead of studying the performance of a specific decoder in a particular example. 

First we simplify the adiabatic noise channel based on physically motivated assumptions. The adiabatic noise channel,
\begin{align}\label{eq_aqc_sectioniii}
    \mathcal{E}_\Gamma(\rho)=\sum_{\gamma\in\Gamma} p_\gamma \mathcal{U}_\gamma \rho \mathcal{U}_\gamma^\dagger,
\end{align}
describes an unknown adiabatic drfit inside the gapped phase, where $\Gamma$ is a set of paths inside a neighborhood $\mathcal{A}$ of $\blambda=0$. However, note that in all examples we studied so far, the image of a code word after an adiabatic drift in a simple neighborhood of the fixed point Hamiltonian depended only on the end point of $\gamma$ rather than the entire path of the adiabatic evolution, except for a global phase which we always ignore. Here we argue that this is more generally true for small simple neighborhoods $\mathcal{A}$ of any code Hamiltonian inside a gapped phase. To show this, we first argue that the adiabatic evolution along any closed path $\gamma_c$ in $\mathcal{A}$ -- that starts and ends at $\blambda=0$ -- implements the identity gate on the codespace of $\mathcal{C}$. This is known to be true in the context of braiding anyons for topological quantum computation, whenever the neighborhood $\mathcal{A}$ is topologically simple\cite{nayak2008non}. In general, an adiabatic evolution along a closed path $\gamma_c$ inside a gapped phase would implement the following unitary in the codespace\cite{wilczek1984appearance} (apart from a global dynamical phase),
\begin{align}\label{eq_closed_path_holonomy}
    &\mathcal{U}_{\gamma_c}=\mathcal{P} \exp(\oint_{\gamma_c} A(\blambda)\cdot\dd \blambda),
\end{align}
where $\mathcal{P}$ is the path ordering operator and $A(\blambda)$ is the Berry connection matrix defined as,
\begin{align}
    A_{ij}(\blambda)=\mel{\varphi_j(\blambda)}{\partial_\blambda}{\varphi_i(\blambda)},
\end{align}
with $\ket{\varphi_i(\blambda)}$ denoting a smooth (in $\blambda$) choice of ground states for $H(\blambda)$.  
As we argue in Appendix \ref{apx_closed_path_holonomy}, for small neighborhoods around $\blambda=0$, one expects the off diagonal terms $A_{ij}$ with $i\ne j$ to vanish, while the diagonal terms $A_{ii}$ to be independent of $i$, up to exponentially small (in system linear size) errors. This follows from local indistinguishability of the ground states inside the gapped phase, plus assuming nearby ground states are related to each other by short local unitary evolution. This in turn shows that $\mathcal{U}_{\gamma_c}$ in Eq. \eqref{eq_closed_path_holonomy} is just a global phase,
\begin{align}
    \mathcal{U}_{\gamma_c}=e^{i\alpha}\mathbb{1}+O(\exp(-L)).
\end{align}
The fact that adiabatic evolution along closed path is proportional to identity, shows that up to a global phase, the image of code states under adiabatic evolution inside a small neighborhood around $\blambda=0$ depends only on the endpoint rather than the entire path. This is because if $\gamma_1$ and $\gamma_2$ are two paths with the same endpoints, then $\mathcal{U}_{\gamma_1} \mathcal{U}_{\gamma_2}^\dagger$ represents an adiabatic evolution along a closed path, and since such a evolution is proportional to identity in the codespace, we have $\mathcal{U}_{\gamma_1}\, \rho\, \mathcal{U}_{\gamma_1}^\dagger=\mathcal{U}_{\gamma_2}\,\rho\, \mathcal{U}_{\gamma_2}^\dagger$. As such, we may express the adiabatic noise channel in Eq. \eqref{eq_aqc_sectioniii} equivalently as,
\begin{align}
\mathcal{E}_\Gamma(\rho)=\sum_{\blambda\in\mathcal{A}} p_\blambda \mathcal{U}_\blambda\,\rho\,\mathcal{U}_\blambda^\dagger,
\end{align}
where $p_\blambda$ is a probability distribution on $\mathcal{A}$ which is related to $p_\gamma$ in Eq. \eqref{eq_aqc_sectioniii} as $p_\blambda=\sum_{\gamma\in\Gamma:\gamma(T)=\blambda}p_\gamma$,  and  $\mathcal{U}_\blambda$ can be taken to be $\mathcal{U}_\gamma$ for any path $\gamma(t)$ with $\gamma(0)=0$ and $\gamma(T)=\blambda$. Accordingly, one may think of $\Gamma$ as just merely defining a probability distribution over $\mathcal{A}$.

\begin{figure}
    \centering
    \includegraphics{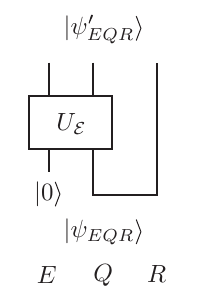}
    \caption{The projection onto the codespace $\rho_Q$ can be purified to a logical Bell state $\ket{\psi_{QR}}$, using a reference system denoted by $R$. The channel $\mathcal{E}$ can also be purified to a unitary $U_\mathcal{E}$, which acts on the system $Q$ plus environment $E$, with the environment initially disentangled from the rest. The existence of a recovery map depends on whether the environment $E$ remains disentangled from the reference $R$ after the application of $U_\mathcal{E}$, i.e. in the output state $\ket{\psi_{EQR}'}$.}
    \label{fig_channelPurification}
\end{figure}

We are going to use the simplified description of the quantum adiabatic channel to understand its reversibility. In general, the reversibility of a quantum channel can be formulated as follows \cite{schumacher2002approximate}.  Let $\mathcal{E}$ be an arbitrary quantum channel. We introduce an environment $E$ and a reference system $R$ to purify the channel and the initial state respectively (see Fig.\ref{fig_channelPurification}). We denote the original system by $Q$. Let $\rho_Q$ denote the projection on the codespace of $\mathcal{C}$ and let $\ket{\psi_{QR}}$ denote its  purification in $QR$, such that $\rho_Q=\tr_R(\ketbra{\psi_{QR}})$. Moreover, we may represent the action of $\mathcal{E}$ on the system $Q$ by a unitary $U_{\mathcal{E}}$ that acts on $Q$ and a non-entangled environment $E$, such that $\mathcal{E}(\rho)=\tr_E(U_\mathcal{E}~\ketbra{0}_E \otimes \rho~ U^\dagger_\mathcal{E})$, for any density matrix $\rho$ in $Q$. We denote the initial pure state of the system plus environment plus reference by $\ket{\psi_{EQR}}=\ket{0}_E\otimes\ket{\psi_{QR}}$ and the final pure state by $\ket{\psi'_{EQR}}=U_\mathcal{E}\ket{\psi_{EQR}}$. Lastly, we denote the reduced density matrices of $\ket{\psi_{EQR}}$ and $\ket{\psi'_{EQR}}$ on a subsystem $A$, by $\rho_A$ and $\rho'_A$ respectively, e.g. $\rho'_{QR}$ corresponds to $\tr_E(\ketbra{\psi'_{EQR}})$. In general, a recovery map $\mathcal{R}$ acting on the system $Q$ exists such that $\mathcal{R}(\rho'_{QR})=\rho_{QR}=\ketbra{\psi_{QR}}$ if and only if $\rho'_{ER}=\rho'_E\otimes \rho'_R$\cite{schumacher1996quantum}. Moreover, such a recovery map, if it exists, would recover any code state as well, i.e. $\mathcal{R}(\mathcal{E}(\ket{\varphi}_Q))=\ket{\varphi}_Q$ for any $\ket{\varphi}_Q\in\mathcal{C}$\cite{schumacher1996quantum}. Furthermore, this statement is robust in the sense that if instead of perfect disentanglement of the environment  and the reference system, $\Delta=\rho'_{ER}-\rho'_E\otimes \rho'_R$ is non-zero, then there exists an approximate recovery map $\mathcal{R}$ acting on $Q$ such that $F(\mathcal{R}(\rho'_{QR}),\rho_{QR})\ge 1-\norm{\Delta}_\text{tr}$, where $F$ stands for the fidelity and $\norm{\Delta}_\text{tr}=\tr[\sqrt{\Delta\,\Delta^\dagger}]$ is the trace norm of $\Delta$\cite{schumacher2002approximate}. This in turn ensures that for any code state $\ket{\varphi}_Q\in\mathcal{C}$ one has, $F(\mathcal{R}(\mathcal{E}(\ket{\varphi}_Q)),\ket{\varphi}_Q)\ge 1-M\norm{\Delta}_\text{tr}$\cite{nielsen2002quantum}, where  $M=\rank(\rho_Q)=2^k$, with $k$ the number of logical qubits encoded in $\mathcal{C}$. Therefore, based on the above discussion, we may argue about the existence of a recovery map for a channel $\mathcal{E}$, by studying $\norm{\Delta}_\text{tr}=\norm{\rho'_{ER}-\rho'_E\otimes\rho'_R}_\text{tr}$.

It is worth noting that while the proof for the existence of the aforementioned recovery maps is constructive\cite{schumacher1996quantum,schumacher2002approximate}, those recovery maps involve complicated many-body non-local measurements and feedback. However, in this section we are only concerned with the existence of a recovery map, not whether it can be implemented  efficiently.

To compute $\Delta$ for the adiabatic noise channel $\mathcal{E}_\Gamma$ we start by the following purification of the channel and the projection onto the code state. Let $\{\ket{\varphi_i}\}_{i=1}^M$ denote a basis set for the codespace of $\mathcal{C}$, and thus $\rho_Q=\frac{1}{M}\sum_{i=1}^M\ketbra{\varphi_i}_Q$. We purify $\rho_Q$ as, 
\begin{align}
    \ket{\psi_{QR}}=\frac{1}{\sqrt{M}}\sum_{i=1}^M \ket{\varphi_i}_Q\ket{i}_R,
\end{align}
where $\{\ket{i}_R\}_{i=1}^M$ represents a set of orthonormal states in $R$. As for the quantum channel $\mathcal{E}_\Gamma$, we purify it via the  unitary $U_{\Gamma}$ which is defined as follows
\begin{align}\label{eq_E_channel}
    U_\Gamma\ket{0}_E\ket{\psi}_Q=\sum_{\blambda} \sqrt{p_\blambda}\ket{\blambda}_E \mathcal{U}_\blambda\ket{\psi}_Q,
\end{align}
where $\{\ket{\blambda}_E\}$ is a set of orthonormal states in $E$ labeled by $\blambda$, and $\ket{\psi}_Q$ is an arbitrary state in $Q$.
Accordingly, the initial pure state of the system plus environment plus reference is given as,
\begin{align}
    \ket{\psi_{EQR}}=\ket{0}_E\otimes\ket{\psi_{QR}}.
\end{align}

and thus the final pure state $\ket{\psi'_{EQR}}$ (see Fig.\ref{fig_channelPurification}) is equal to,
\begin{align}
    \ket{\psi'_{EQR}}&=\frac{1}{\sqrt{M}}\sum_{i=1}^M \sum_{\blambda}\sqrt{p_\blambda}\ket{\blambda}_E \mathcal{U}_\blambda \ket{\varphi_i}_Q \ket{i}_R\\
    &=\frac{1}{\sqrt{M}}\sum_{i=1}^M \sum_{\blambda}\sqrt{p_\blambda}\ket{\blambda}_E \ket{\varphi_{i}(\blambda)}_Q \ket{i}_R,
\end{align}
where we have defined,
\begin{align}
    \ket{\varphi_{i}(\blambda)}_Q=\mathcal{U}_\blambda\ket{\varphi_i}_Q.
\end{align}

Based on the above discussion, we set out to compute $\Delta=\rho'_{ER}-\rho'_{E}\otimes \rho'_R$. Note that,
\begin{align}
    &\ketbra{\psi'_{EQR}}=\nonumber \\
    &\frac{1}{M}\sum_{i,i'}\sum_{\blambda,\blambda'}\sqrt{p_\blambda p_{\blambda'}}\ketbra{\blambda,\varphi_i(\blambda),i}{\blambda',\varphi_{i'}(\blambda'),i'},
\end{align}
and hence,
\begin{align}
    &\rho'_{ER}=\nonumber\\
    &\frac{1}{M}\sum_{i,i'}\sum_{\blambda,\blambda'}\sqrt{p_\blambda p_{\blambda'}}\braket{\varphi_{i'}(\blambda')}{\varphi_{i}(\blambda)}_Q\ketbra{\blambda}{\blambda'}_E \otimes\ketbra{i}{i'}_R,
\end{align}
which yields, 
\begin{align}
    &\rho'_R=\frac{1}{M}\sum_{i}\ketbra{i}_R,\\
    &\rho'_E=\sum_{\blambda,\blambda'}\sqrt{p_\blambda p_{\blambda'}} \qty(\frac{1}{M}\sum_{i}\braket{\varphi_{i}(\blambda')}{\phi_{i}(\blambda)}_Q)\ketbra{\blambda}{\blambda'}_E,
\end{align}
where in the first line we have used $\braket{\varphi_{i}(\blambda)}{\varphi_{i'}(\blambda)}_Q=\delta_{i,i'}$, and $\sum_\blambda p_\blambda=1$. Hence we find,
\begin{align}\label{eq_delta_mels}
    \Delta=&\rho'_{ER}-\rho'_E\otimes\rho'_R\nonumber\\
        =&\frac{1}{M}\sum_{i\ne i'}\sum_{\blambda,\blambda'}\sqrt{p_\blambda p_{\blambda'}}\braket{\varphi_{i'}(\blambda')}{\varphi_{i}(\blambda)}_Q\ketbra{\blambda}{\blambda'}_E \otimes\ketbra{i}{i'}_R\nonumber\\
        &+\frac{1}{M}\sum_i \sum_{\blambda,\blambda'}\sqrt{p_\blambda p_{\blambda'}}\sigma_{i}(\blambda',\blambda) \ketbra{\blambda}{\blambda'}_E\otimes \ketbra{i}{i}_R,
\end{align}
where
\begin{align}
    \sigma_{i}(\blambda',\blambda)=\braket{\varphi_{i}(\blambda')}{\varphi_{i}(\blambda)}_Q-\frac{1}{M}\sum_{j=1}^M \braket{\varphi_{j}(\blambda')}{\varphi_{j}(\blambda)}_Q.
\end{align}
To proceed further we need to be more specific about the set $\Gamma$, or equivalently the probability distribution $p_\blambda$. In what follows we consider the case where $p_\blambda$ is non-zero on a discrete set of $K$ points in the gapped phase $\{\blambda_\mu\}_{\mu=0}^K$, that correspond to the set of end points of the paths in $\Gamma$. We are interested in taking $K$ to infinity to cover the whole neighborhood $\mathcal{A}$. However, since we are taking the thermodynamic limit $N\to\infty$ as well, the order of the two limits could in principle matter. As we argue in the next sections, if the thermodynamic limit is taken first, in general the orthogonality catastrophe alone would ensure the existence of a recovery map. However, in the case where $K\to \infty$ limit is taken before the thermodynamic limit, we find certain conditions by the ground states, which are sufficient for the existence of a recovery map. It is worth noting that in practice, it is the latter order of limits which is of interest. 

\emph{Taking the thermodynamic limit first:} We now consider taking the thermodynamic limit first. For generic perturbations, one expects the many-body ground states at two different points of the phase diagram to be orthogonal in the thermodynamic limit, i.e. for fixed $|\blambda-\blambda'|>0$ 
\begin{align}\label{eq_ortho_catastrophe}
    \lim_{N\to\infty} \braket{\varphi_{i}(\blambda)}{\varphi_{i'}(\blambda')}=0.
\end{align}
While this limit can be easily verified for the ground states of the transverse field Ising model from the analytical expressions of the ground states(see Appendix \ref{apx_JWandFFM}), one expects it to hold more generally due to the Anderson orthogonality catastrophe\cite{anderson1967infrared}. Moreover, note that the $\blambda=\blambda'$ matrix elements of the $\Delta$ matrix in Eq. \eqref{eq_delta_mels} are trivially $0$. Therefore, in the thermodynamic limit all elements of the $\Delta$ matrix go to zero. Since we are taking $K$ (and $M$) to be fixed, the dimension of the the $\Delta$ matrix, i.e. $KM$ is also fixed. As such, $\lim_{N\to\infty} \norm{ \Delta}_\text{tr}=0$, which means perfect recovery is possible in the thermodynamic limit. 

$K$ is assumed to be fixed in the above analysis, but nonetheless, the same line of argument holds if one takes $K\to \infty$ limit at the same time as the thermodynamic limit, given that $K$ grows not so fast with $N$. For example, in the transverse field Ising model the overlap decays exponentially with the system size (see Appendix \ref{apx_JWandFFM}) in the following form,
\begin{align}
    \braket{\varphi_\pm(g)}{\varphi_\pm(g')}=O\Big(\exp(-\frac{(g-g')^2}{32} N)\Big),
\end{align}
when $|g-g'|\ll 1$. Accordingly, if one takes $K$ to infinity such that the spacing between possible values of $g$ scales as $\sim N^{-1/4}$, each matrix element in $\Delta$ vanishes at least exponentially in $\sqrt{N}$, while dimension of $\Delta$, i.e. $KM$, grows only polynomially in $N$, thus ensuring $\lim_{K,N\to\infty} \norm{\Delta}_\text{tr}=0$. 

It is worth noting that the perfect recovery in this case is mostly about the infinite size of the Hilbert space rather than the physics of adiabatic evolution or phases of matter. Indeed, for a fixed $K$, perfect recovery is possible in the thermodynamic limit even if  $\mathcal{U}_\blambda$ unitaries in Eq. \eqref{eq_E_channel} are chosen at random via Haar measure on $U(N)$. 

\emph{Taking the $K\to\infty$ limit first:} We now consider taking $K\to\infty$ limit first and then taking the thermodynamic limit, which is of more relevance than the other order of limits. To upper bound  $\norm{\Delta}_{\tr}=\tr\sqrt{\Delta \Delta^\dagger}$, which controls the fidelity of the approximate recovery map\cite{schumacher2002approximate}, we compute the Frobenius norm $\norm{\Delta}_F=\sqrt{\tr(\Delta \Delta^\dagger)}$ first, and then use the following inequality
\begin{align}\label{eq_tr_for_ineq}
    \norm{\Delta}_{\text{tr}}^2 \le \text{rank}{\Delta}\times  \norm{\Delta}_F^2.
\end{align}
Note that $\rank(\rho'_R)=M$, and since $\ket{\psi'_{EQR}}$ is pure,  $\rank(\rho'_{ER})\le 2^N$ and $\rank(\rho'_{E})\le M\,2^N$. Therefore, due to subaditivity of rank, we have 
\begin{align}\label{eq_rank_upbd}
\rank(\Delta)\le (1+M^2)2^N.    
\end{align}
According to Eq. \eqref{eq_delta_mels} we have,
\begin{align}\label{eq_frb_norm_Delta}
    \norm{\Delta}_F^2=\frac{1}{M^2}\sum_{\blambda,\blambda'}p_\blambda p_{\blambda'}\Bigg[ 
    &\sum_{i\ne i'} |\braket{\varphi_{i'}(\blambda')}{\varphi_{i}(\blambda)}_Q|^2\nonumber\\
    &+ \sum_{i} |\sigma_i(\blambda,\blambda')|^2\Bigg]
\end{align}
Both terms inside the square brackets vanish for $\blambda'=\blambda$. If they remain exponentially small for $\blambda\ne \blambda'$, we may use Eq. \eqref{eq_tr_for_ineq} and Eq. \eqref{eq_rank_upbd} to show that the trace norm of $\Delta$ vanishes in the thermodynamic limit. Therefore, it gives a sufficient condition for perfect recovery. More concretely, assume there exist a neighborhood $\mathcal{A}$ of the fixed point $\blambda=0$ and positive numbers $c,c'<1/\sqrt{2}$ and $N_0>0$, such that for $N \ge N_0$ and any $\blambda,\blambda' \in \mathcal{A}$ and $i'\ne i$,
\begin{align}
    &\Big|\braket{\varphi_{i'}(\blambda')}{\varphi_{i}(\blambda)}_Q\Big| < b~ c^{N}\label{eq_cond_1},\\
    &\Big|\braket{\varphi_{i}(\blambda')}{\varphi_{i}(\blambda)}_Q- \braket{\varphi_{i'}(\blambda')}{\varphi_{i'}(\blambda)}_Q|<b'~c'^{N} \label{eq_cond_2},
\end{align}
for some constants $b,b'$. Note that Eq. \eqref{eq_cond_2} ensures $|\sigma_i(\blambda,\blambda')|<b'\,c'^N$. If these assumptions hold, by combining Eq. \eqref{eq_tr_for_ineq}, Eq. \eqref{eq_rank_upbd} and Eq. \eqref{eq_frb_norm_Delta}  we find that $\norm{\Delta}_{\tr} = O((2c^2)^N)+O((2c'^2)^N)$ which goes to zero as one takes the $N\to\infty$ limit, and hence ensuring perfect recovery in the thermodynamic limit. Eq. \eqref{eq_cond_1} and Eq. \eqref{eq_cond_2} may be viewed as analogous to Knill-Laflamme conditions\cite{knill1997theory} for approximate quantum error correction\cite{beny2010general,faist2020continuous}. 

\subsection{Phases Fulfilling  Conditions for  Recovery}
It is instructive to revisit the examples we considered so far 
in light of the above sufficient conditions. We first remark that in all examples we have considered,  Eq. \eqref{eq_cond_1} is trivially satisfied because the corresponding wave function overlaps are identically zero due symmetry arguments. In the example of the rotated Ising model (Section \ref{sec_ising_rotation}) it is easy to compute the overlaps appearing in Eq. \eqref{eq_cond_2},
\begin{align}
    \Big|\braket{\varphi_{+}(\theta')}{\varphi_{+}(\theta)}-\braket{\varphi_{-}(\theta')}{\varphi_{-}(\theta)}\Big|=2 \sin^N \frac{\theta'-\theta}{2},
\end{align}
so Eq. \eqref{eq_cond_2} is satisfied in $\theta \in (-\pi/4,\pi/4)$ neighborhood of $0$. This is obviously not a tight bound (mainly because the inequality \eqref{eq_tr_for_ineq} is not tight), given that we know perfect recovery is possible  for $\theta\in (-\pi/2,\pi/2)$. As for the transverse field Ising model, perturbative arguments based on local indistinguishably of different ground states  (see Appendix \ref{apx_perturb_theory_TFIM}) show that\footnote{One can use the analytic solution to show that $\frac{\braket{\varphi_{+}(g')}{\varphi_{+}(g)}}{\braket{\varphi_{-}(g')}{\varphi_{-}(g)}}=1+\varepsilon(N)$, with $\varepsilon(N)$ vanishing in $N$ faster than any polynomial (see Appendix \ref{apx_JWandFFM})} 
\begin{align}
    \braket{\varphi_{+}(g')}{\varphi_{+}(g)}=\braket{\varphi_{-}(g')}{\varphi_{-}(g)}=\mathcal{O}(g^N)+\mathcal{O}(g'^N).
\end{align}
Hence, Eq. \eqref{eq_cond_2} is satisfied if $g$ and $g'$ are restricted to a small enough neighborhood around $0$. Again, this is not tight, because we already know that perfect recovery is possible for $g,g'<1$. 

More generally, in a generic spontaneous symmetry broken phase protected by a global symmetry $G$, one could always choose the states $\ket{i}_Q$ to be eigenstates of $G$ with different charges. This will ensure that \eqref{eq_cond_1} is trivially satisfied throughout the phase. Furthermore, \eqref{eq_cond_2} follows from perturbative arguments similar to that of the TFIC: Since $G$ is a global symmetry, $|\text{Supp}(G)|=\mathcal{O}(N)$ and thus it is only at the $\mathcal{O}(N)$-th order of the perturbation theory that a local symmetric perturbation can tell the difference between different symmetry sectors. Thus, perfect recovery from an adiabatic drift should be possible at least in a neighborhood of the fixed point Hamiltonian of a spontaneous symmetry broken phase with a global symmetry. The same statement should also hold for $0$-form symmetry protected topological phases
based on the same line of reasoning (see also Appendix \ref{apx_clustermodel}). 

Lastly, while similar arguments can be used for long range entangled topological phases of matter in $D$ dimensions, e.g. the perturbed toric code example studied in Section \ref{sec_toric_code}, it only results in upper bounds which are exponentially small in the linear size of the system, i.e. $L=N^{1/D}$ instead of $N$, which is not fast enough to satisfy \eqref{eq_cond_1} and \eqref{eq_cond_2}. However the bounds in \eqref{eq_cond_1} and \eqref{eq_cond_2} are just sufficient conditions but not necessary, and it could be the case that in general an exponential suppression in linear size of the system is good enough for a successful recovery, as is the case in the perturbed toric code example considered in Section \ref{sec_toric_code}. In particular, we expect the arguments in Section \ref{sec_ising_and_toric_as_channels} hold more generally for any topological quantum error correcting code, namely that measuring stabilizers would decohere the adiabatic noise channel into an effective correlated Pauli noise channel, with a correlation length that vanishes at $\blambda=0$ and diverges at the phase boundary. As such, one would expect that MWPM decoder is able to reverse the adiabatic noise channel in a small neighborhood of the fixed point, while it fails beyond an error correction threshold inside the gapped phase. 

\textit{Note Added:}  During the completion of this work, another work appeared which studies the decoding problem for quantum information encoded in the ground-space of perturbed stabilizer codes \cite{zhong2024advantage}.

\begin{acknowledgments}
AL and SV acknowledge useful discussions with Utkarsh Agrawal, Victor V. Albert, Yimu Bao, Shankar Balasubramanian, Matthew P. A. Fisher, Ethan Lake, and Yaodong Li.  SV acknowledges the support of the Alfred P. Sloan Foundation through a Sloan Research Fellowship. Use was made of computational facilities purchased with funds from the National Science Foundation (CNS-1725797) and administered by the Center for Scientific Computing (CSC). The CSC is supported by the California NanoSystems Institute and the Materials Research Science and Engineering Center (MRSEC; NSF DMR 2308708) at UC Santa Barbara. 
\end{acknowledgments}

\bibliographystyle{plain}
\bibliography{refs}


\clearpage
\newpage
\appendix
\section{Adiabatic Drift for the TFIC \label{apx_JWandFFM}}

This section contains the detailed free fermion calculation related to the ground states of the $1$d transverse field Ising model.
\subsection{Jordan-Wigner transformation}
 We use the following Jordan-Wigner transformation to write the Hamiltonian \eqref{eq:orginial_hamiltonina} in terms of Majoranas,
 \begin{align}
     &\gamma_i =\prod_{j<i} X_j \cdot Z_i\\
     &\gammatilde_i =\prod_{j<i} X_j \cdot Y_i.
 \end{align}
 Accordingly, we have,
 \begin{align}
     X_i&=i\gamma_i \gammatilde_i\\
     Z_i Z_{i+1}&=i \gammatilde_i \gamma_{i+1},
 \end{align}
For convenience we are also going to use a unified notation for $\gamma$ and $\gammatilde$ Majoranas and denote both of them by $a_i$ for $i=1,\cdots,2N$:
\begin{align}
    &a_{2i-1}\equiv \gamma_i\\
    &a_{2i}\equiv \gammatilde_i.
\end{align}
We also denote the collection of all $a$ Majorana modes by $\ba=(a_1,\cdots, a_{2N})$.
 \subsection{Open Boundary Conditions}
 Under JW transformation the Hamiltonian \eqref{eq:orginial_hamiltonina} becomes,
 \begin{align}\label{eq:majorana_hamiltonian}
     H=-i \sum_{i=1}^{N-1}\gammatilde_i \gamma_{i+1}-ig\sum_{i=1}^N \gamma_i \gammatilde_{i}.
 \end{align}
 
\subsection{Closed Boundary Conditions}
If the chain is on the closed boundary, one needs to be more careful with the Jordan-Wigner transformation. Notably, the spin-spin interaction term around the boundary would be mapped to,
\begin{align}
    Z_NZ_1&=(Z_1Z_2)(Z_2Z_3)\cdots(Z_{N-1}Z_N)\\
            &=i^{N-1} \gammatilde_1\gamma_2 \gammatilde_2 \gamma_3\cdots \gammatilde_{N-1}\gamma_N\\
    &=-i\gamma_1 P \gammatilde_N=-iP\gammatilde_N\gamma_1.
\end{align}
where $P=\prod_i X_i=i^N\prod_i(\gamma_i\gammatilde_i)$ is the parity operator. Since $P$ commutes with the Hamiltonian, we may restrict the Hamiltonian to $P=\pm 1$ subspaces and diagonalize each subspace separately. For the $P=-1$ subspace, we recover the Kitaev model on a closed chain, while for the $P=1$, the pairing term across the boundary acquires a minus sign, which can be thought of as twisted boundary conditions or flux insertions.

When looking at the fermionic Hamiltonian for the $P=1$ ($P=-1$) sector, only the states with even (odd) parity are physical. One thing to note is that $P$ is not invariant under general orthogonal transformation of Majoranas. In general if $\ba'=Q \ba$, we have,
\begin{align}
    P(\ba')=\det(Q)P(\ba).
\end{align}

Define the following fermionic modes:
\begin{align}
    f_i&=\frac{\gamma_i+i\gammatilde_i}{2}\\
    f^\dagger_i&=\frac{\gamma_i-i\gammatilde_i}{2},
\end{align}
with the inverse transformation
\begin{align}
    &\gamma_i=f_i + f_i^\dagger\\
    &\gammatilde_i=i(f_i^\dagger-f_i)
\end{align}
The Hamiltonian in terms of $f$ fermions reads as,
\begin{align}\label{eq_f_with_P}
    H=&\sum_{i=1}^{N-1}(f_i^\dagger-f_i)(f_{i+1}+f_{i+1}^\dagger)-P (f_N^\dagger-f_N)(f_{1}+f_{1}^\dagger)\nonumber\\
    &+g\sum_{i=1}^N (f_i+f_i^\dagger)(f_i^\dagger-f_i)\nonumber\\
    =&\sum_{i=1}^{N-1}f_i^\dagger f_{i+1}+f_i^\dagger f_{i+1}^\dagger+h.c.-P (f_N^\dagger f_{1}+f_N^\dagger f_{1}^\dagger) + h.c.\nonumber\\
    &-2g\sum_{i=1}^N f_i^\dagger f_i+\text{const.}
\end{align}
where $h.c.$ stands for hermitian conjugate. For simplicity, in the rest of the calculation we will drop the constants in  Hamiltonians without notice.

Let $-P=e^{i\phi}$, with $\phi=\pi$ for even parity and $\phi=0$ for odd parity. 
We define the following momentum space modes (assuming $N$ is even),
\begin{align}\label{eq_f_k_def}
    f_k\equiv\frac{1}{\sqrt{N}}\sum_{j=1}^N e^{-ikj}f_j,
\end{align}
for 
\begin{align}\label{eq_possible_k}
     k=\frac{2\pi (m+\phi/2\pi)}{N},~m=-\frac{N}{2},-\frac{N}{2}+1,\cdots,\frac{N}{2}-1,
\end{align}
which results in $-\pi +\frac{\phi}{N}\le k \le \pi -\frac{2\pi}{N}+\frac{\phi}{N}$. Note that $f_k=f_{k+2\pi}$. We also have the reverse transformation
\begin{align}
    f_j=\frac{1}{\sqrt{N}}\sum_{k} e^{+ikj}f_k,\label{eq:invF}
\end{align}
Eq. \eqref{eq:invF} holds as an equality for $j=1,\cdots,N$. Moreover, we use Eq. \eqref{eq:invF} to define $f_j$ for $j>N$. In particular we have,
\begin{align}
    &f_{j+N}=\frac{1}{\sqrt{N}}\sum_{k} e^{+ikj}e^{+ikN}f_k=e^{i\phi}f_j=-P\, f_j\\
    &f^\dagger_{j+N}=e^{-i\phi}f^\dagger_j=-P\, f^\dagger_j,
\end{align}
where we have used the fact that $e^{-i\phi}=e^{i\phi}$. Therefore, we may write the Hamiltonian Eq. \eqref{eq_f_with_P} simply as,
\begin{align}\label{eq_H_witout_P}
    H=&\sum_{j=1}^{N}f_j^\dagger f_{j+1}+f_j^\dagger f_{j+1}^\dagger+h.c.-2g\sum_{j=1}^N f_j^\dagger f_j+const.
\end{align}
For future references, the following sum identities holds irrespective of $\phi$,
\begin{align}
    &\sum_{j}e^{i(k-k')j}=N\delta_{k,k'},\\
    &\sum_j e^{i(k+k')j}=N\delta_{k,-k'}\\
    &\sum_{k}e^{i(j-j')k}=N\delta_{j,j'}.
\end{align}
By using the above relations, one can re-write the Hamiltonian \eqref{eq_H_witout_P} in the momentum space as,
\begin{align}
     H=&2\sum_{k} \Big(\cos(k)-g\Big) f^\dagger_k f_k \nonumber \\ 
     &+2i\sum_{k>0}\sin(k)\Big( f^\dagger_k f^\dagger_{-k}+f_{k} f_{-k}\Big)\nonumber \\
     =&2\sum_{0<k<\pi}
     \begin{pmatrix}
        f^\dagger_k & f_{-k}
    \end{pmatrix}
    \begin{pmatrix}
        \cos(k)-g & i \sin(k)\\
        -i \sin(k) & -\Big(\cos(k)-g\Big)
    \end{pmatrix}
    \begin{pmatrix}
        f_k\\
        f^\dagger_{-k}
    \end{pmatrix}\nonumber\\
    &+\delta_{\phi,0}\Bigg[ 2(1-g)f^\dagger_{k=0}f_{k=0}+2(-1-g)f^\dagger_{k=-\pi}f_{k=-\pi}\Bigg]\nonumber\\
    =&\sum_{0<k<\pi}
     \begin{pmatrix}
        f^\dagger_k & f_{-k}
    \end{pmatrix}
    \bmh_k
    \begin{pmatrix}
        f_k\\
        f^\dagger_{-k}
    \end{pmatrix}\nonumber \\
    &+\delta_{\phi,0}\Bigg[ 2(1-g)f^\dagger_{k=0}f_{k=0}+2(-1-g)f^\dagger_{k=-\pi}f_{k=-\pi}\Bigg].
 \end{align}
 with
 \begin{align}
     \bmh_k=2(\cos(k)-g)\sigma_z-2\sin(k)\sigma_y
 \end{align}
 where $\sigma_z$ and $\sigma_y$ are Pauli-$Z$ and Pauli-$Y$ matrices. 
Define $\epsilon_k$ and $-\pi\le\theta_k<\pi$ as follows,
 \begin{align}
     &\epsilon_k=2\sqrt{1+g^2-2g\cos(k)}\label{eq:epsilonk}\\
     &\sin(\theta_k)=\frac{2\sin(k)}{\epsilon_k},\quad \cos(\theta_k)=\frac{2(\cos(k)-g)}{\epsilon_k}\label{eq:theta_k},
 \end{align}
 so that $\bmh_k$ acquires the following form,
 \begin{align}
     \bmh_k=\epsilon_k\Big[\cos(\theta_k)\sigma_z-\sin(\theta_k)\sigma_y\Big].
 \end{align}
The following unitary diagonalizes $\bmh_k$,
 \begin{align}
     &U_k=
     \begin{pmatrix}
        \cos(\theta_k/2)&i\sin(\theta_k/2)\\
        i\sin(\theta_k/2)&\cos(\theta_k/2)
    \end{pmatrix},\\
    & \bmh_k =
    U_k^\dagger\begin{pmatrix}
        \epsilon_k&0\\
        0&-\epsilon_k
    \end{pmatrix}U_k.
 \end{align}
 Hence, if we define $(b_k~b^\dagger_{-k})$ modes for $0<k<\pi$ as the following,
 \begin{align}
     \begin{pmatrix}
        b_k\\
        b_{-k}^\dagger
    \end{pmatrix}
    &=U_k
    \begin{pmatrix}
        f_k\\
        f_{-k}^\dagger
    \end{pmatrix}\nonumber \\
    &=
    \begin{pmatrix}
        \cos(\theta_k/2) f_k+i\sin(\theta_k/2) f^\dagger_{-k}\\
        i\sin(\theta_k/2) f_k+\cos(\theta_k/2) f^\dagger_{-k},
    \end{pmatrix}\label{eq_b_def_two}
 \end{align}
 it is easy to check that $b_k$ operators satisfy the correct commutation relations and that the Hamiltonian takes the following diagonal form,
 \begin{align}
 H=&\sum_{k>0}\epsilon_k(b_k^\dagger b_k+b_{-k}^\dagger b_{-k} )\nonumber \\
    &+\delta_{\phi,0}\Bigg[ 2(1-g)f^\dagger_{k=0}f_{k=0}+2(-1-g)f^\dagger_{k=-\pi}f_{k=-\pi}\Bigg].
 \end{align}
 Note that if we use Eq. \eqref{eq:theta_k} to define $\theta_k$ for $k<0$ as well, we find that,
 \begin{align}
     \theta_{-k}=-\theta_k.
 \end{align}
Hence, instead of the two equations in \eqref{eq_b_def_two}, we can define $b_k$ for any $k\ne 0,\pi$ via the following single expression,
 \begin{align}
     b_k=\cos(\theta_k/2) f_k+i\sin(\theta_k/2) f^\dagger_{-k},\label{eq:bk}
 \end{align}
Moreover, when $-1<b<1$, if we use Eq. \eqref{eq:theta_k} formally to find $\theta_{0}$ and $\theta_{-\pi}$, we find 
 \begin{align}
     \theta_{0}=0, \quad \theta_{-\pi}=-\pi.
 \end{align}
 Thus if we use Eq. \eqref{eq:bk} formally to define $b_0$ and $b_{-\pi}$, we find,
 \begin{align}
     b_0=f_{k=0},\quad b_{-\pi}=-if^\dagger_{k=\pi}=-if^\dagger_{k=-\pi}.\label{eq:b0pi}
 \end{align}
 Therefore, for $-1<b<1$, we may write the Hamiltonian as simply,
 \begin{align}\label{eq_hamiltonian_diag_bk}
     H=\sum_{-\pi\le k<\pi} \epsilon_k b_k^\dagger b_k,
 \end{align}
 where $\epsilon_k$ is given by Eq. \eqref{eq:epsilonk}, $b_k$ is given by Eq. \eqref{eq:bk} and $\theta_k$ is given by Eq. \eqref{eq:theta_k} for all $-\pi\le k<\pi $. The difference between $\phi=0,\pi$ is now just in the valid values that $k$ assumes. 

\subsubsection{Parity in the canonical frame}
 Define the matrix $w$ to be
 \begin{align}
 w=
          \begin{pmatrix}
         1&1\\
         -i&i
     \end{pmatrix}.
 \end{align}
 Then the transformation from $f_k$ complex fermions to $\gamma_k$ and $\gammatilde_k$ Majorana fermions can be written as, 
 \begin{align}
     \begin{pmatrix}
         \gamma_k\\
         \gammatilde_k\\
         \gamma_{-k}\\
         \gammatilde_{-k}
     \end{pmatrix}
     =
     \begin{pmatrix}
         w&\mathbf{0}\\
         \mathbf{0}&w
     \end{pmatrix}
     \begin{pmatrix}
         f_k\\
         f^\dagger_k\\
         f_{-k}\\
         f^\dagger_{-k}.
     \end{pmatrix}
     =W
    \begin{pmatrix}
         f_k\\
         f^\dagger_k\\
         f_{-k}\\
         f^\dagger_{-k}.
     \end{pmatrix}.
 \end{align}
 Let $\mathcal{U}_k$ be the matrix that transforms $f_k$ to $b_k$, i.e.
 \begin{align}
          \begin{pmatrix}
         b_k\\
         b^\dagger_k\\
         b_{-k}\\
         b^\dagger_{-k}.
     \end{pmatrix}
     =
     \mathcal{U}_k
               \begin{pmatrix}
         f_k\\
         f^\dagger_k\\
         f_{-k}\\
         f^\dagger_{-k}.
     \end{pmatrix}.
 \end{align}
 Therefore, the transformation $\mathcal{U}_k$ in the complex fermions, induces the transformation $W\mathcal{U}_k W^{-1}$ in the Majorana fermions. Hence, the parity under this transformation acquires a sign given by $\det(W\mathcal{U}_k W^{-1})=\det(\mathcal{U}_k) $. On the other hand $\mathcal{U}$ can be written as,
 \begin{align}
     \mathcal{U}_k= O^T 
     \begin{pmatrix}
         U_k&\mathbf{0}\\
         \mathbf{0}&U_{-k}
     \end{pmatrix}
     O,
 \end{align}
 where $O$ is the orthogonal matrix that exchanges the second and forth columns and rows,
 \begin{align}
     O=
     \begin{pmatrix}
         1&0&0&0\\
         0&0&0&1\\
         0&0&1&0\\
         0&1&0&0\\
     \end{pmatrix}.
 \end{align}
 Therefore, 
 \begin{align}
     \det(\mathcal{U}_k)=\det(U_k)\det(U_{-k})=1.
 \end{align}
 For $\phi=\pi$ (i.e. $P=1$), all transformations are governed by $U_k$ rotations, so the parity in the $b_k$ fermions is the same as the parity in the $f_k$ fermions, which is the same as the parity in the $f_j$ fermions (because the Fourier transformation conserves particle number, it trivially conserves parity). On the other hand, for $\phi=0$ (i.e. $P=-1$), the transfromation of $f_{k=0}$ and $f_{k=-\pi}$ to $b_0$ and $b_{-\pi}$ does not follow the $U_k$ form, but follows from Eq. \eqref{eq:b0pi}. The transformation for $f_0$ is identity while the transformation for $f_{-\pi}$ is particle-hole tranformation which flips the parity. Therefore, the odd parity sector of $f_j$ fermions corresponds to the even parity sector of $b_k$ fermions for $\phi=0$. This shows that the desired states in both $P=1$ and $P=-1$ sectors  are the even sectors in $b_k$ fermions. In particular, the odd and even parity (in $f$) ground states $\ket{g(b)\pm}$ correspond to ground states of Hamiltonian \eqref{eq_hamiltonian_diag_bk} for $\phi=0,\pi$, which have no $b_k$ fermions (and thus have even parity in $b_k$ fermions).

\subsubsection{The perturbed ground states}
 Let $U'_k$ be $U_k$  at $g=0$. Note that at $b=0$, we have $\theta_k=k$,
 \begin{align}
     U'_k=
     \begin{pmatrix}
        \cos(k/2)&i\sin(k/2)\\
        i\sin(k/2)&\cos(k/2)
    \end{pmatrix}
 \end{align}
 Define the $c_k$ fermionic operators to be the eigen modes at $g=0$, hence:
 \begin{align}\label{eq_ck_anihilates_ghz}
     c_k \ket{\ghz+}=0,
 \end{align}
 for all $k$. And the same is true for $\ket{\ghz -}$ and $c_k$ operators in the odd parity subspace. We may simply write $b_k$ operatros in terms of $c_k$ operators,
 \begin{align}
     \begin{pmatrix}
        b_k\\
        b_{-k}^\dagger
    \end{pmatrix}
    =U_k (U'_k)^\dagger
    \begin{pmatrix}
        c_k\\
        c_{-k}^\dagger
    \end{pmatrix}
    =\widetilde{U}_k 
    \begin{pmatrix}
        c_k\\
        c_{-k}^\dagger
    \end{pmatrix},
 \end{align}
with
\begin{align}
         \widetilde{U}_k=
     \begin{pmatrix}
        \cos(\frac{\theta_k-k}{2})&i\sin(\frac{\theta_k-k}{2})\\
        i\sin(\frac{\theta_k-k}{2})&\cos(\frac{\theta_k-k}{2})
    \end{pmatrix}
\end{align}
 For Convenience, we define $\phi_k=\theta_k-k$. Hence we have,
 \begin{align}
     b_k=\cos(\phi_k/2)\,c_k +i\sin(\phi_k/2)\, c^\dagger_{-k}.
 \end{align}
 Note that for $k=0$ and $k=-\pi$ (and only for theses values) we have $\phi_0=\phi_{-\pi}=0$, and hence
 \begin{align}
     b_0=c_0,\qquad b_{-\pi}=c_{-\pi}.
 \end{align}
 As such, we may write the symmetric ad anti-symmetric ground states $\ket{\varphi_\pm(g)}$ as,
 \begin{align}
    \ket{\varphi_\pm(g)}=\mathcal{N}\prod_{k\ne 0,-\pi} b_k \ket{\ghz_\pm}.
 \end{align}
 where $\mathcal{N}$ is a normalization constant with a arbitrary phase. Note that this state would be annihilated by all $b_k$ operators.  By expanding $b_k b_{-k}$ in terms of $c_k$ operators and using Eq. \eqref{eq_ck_anihilates_ghz}, we may arrive at the following expressions for the ground states in the even and odd parity sectors,
 \begin{align}\label{eq_gs_analytic_solution}
     \ket{\varphi_\pm(g)}=\prod_{0\le k \le \pi} (\cos(\phi_k/2)+i\sin(\phi_k/2)c^\dagger_k c^\dagger_{-k})\ket{\ghz_\pm},
 \end{align}
 up to an arbitrary global phase. Note that these states are normalized. 

 Now we argue that Eq. \eqref{eq_phase_fixing} fixes the global phase in Eq. \eqref{eq_gs_analytic_solution} to be 1. In particular, note that
\begin{align}
    &\frac{\dd }{\dd g}\ket{\varphi_\pm(g)}=\nonumber\\
    &\sum_{0\le q\le \pi}\frac{1}{2}\Big(-\sin(\phi_q/2)+i \cos(\phi_q/2)c^\dagger_qc^\dagger_{-q}\Big)\frac{\dd \phi_q}{\dd g}\nonumber\\
    &\times\prod_{\substack{0\le k \le \pi\\
                  k\ne q}}
    (\cos(\phi_k/2)+i\sin(\phi_k/2)c^\dagger_k c^\dagger_{-k})\ket{\ghz_\pm}.
\end{align}
However, since
\begin{align}
    \bra{\ghz_\pm}&(\cos(\phi_q/2)-i\sin(\phi_q/2)c_{-q} c_{q})\nonumber\\
    &\times (-\sin(\phi_q/2)+i \cos(\phi_q/2)c^\dagger_qc^\dagger_{-q})\ket{\ghz_\pm}=0
\end{align}
we have
\begin{align}
    \bra{\varphi_\pm(g)}\frac{\dd}{\dd g}\ket{\varphi_\pm(g)}=0.
\end{align}
This choice of gauge is equivalent to fixing $\braket{\ghz_\pm}{\varphi_\pm(g)}$ to be real, which is the way we have fixed the phase in our numerics.

\subsubsection{Computing the overlap $\braket{\varphi_\pm(g)}{\varphi_\pm(g')}$}
We can use the analytic expression for the ground states in to compute the overlap between the ground states at different $g$ and $g'$. Note that the ground states at different transverse fields $g$ and $g'$ can be related to each other as,
 \begin{align}\label{eq_gs_b_and_bprim}
     \ket{\varphi_\pm(g')}=\prod_{0\le k \le \pi} (\cos(\phi_k/2)+i\sin(\phi_k/2) {\widetilde{c}}^\dagger_k {\widetilde{c}}^\dagger_{-k})\ket{\varphi_\pm(g)},
 \end{align}
where $\ket{\varphi_\pm(g)}$ and $\widetilde{c}_k$ are the ground states and the eigen-modes at $g$ (hence $\widetilde{c}_k\ket{\varphi_\pm(g)}=0$ for all $k$). Moreover, in an abuse of notation, $\phi_k=\theta_k(g')-\theta_k(g)$, where $\theta_k(g)$ is defined via Eqs.\eqref{eq:theta_k}.  The parameter $k$ takes the following values for $\pm$ parities,
 \begin{align}\label{eq_k_values}
     k=
    \begin{cases}
    \frac{2\pi m}{N}+\frac{\pi}{N}   & +\\
    \frac{2\pi m}{N}                 & -
    \end{cases}
     ,\quad m=-\frac{N}{2},-\frac{N}{2}+1,\cdots,\frac{N}{2}-1,
\end{align}
which is basically Eq. \eqref{eq_possible_k} but written explicitly for each parity sector. The expression in Eq. \eqref{eq_gs_b_and_bprim} follows from the same line of argument that lead to Eq. \eqref{eq_gs_analytic_solution}. 

\begin{figure}
    \centering
    \includegraphics[width=\columnwidth]{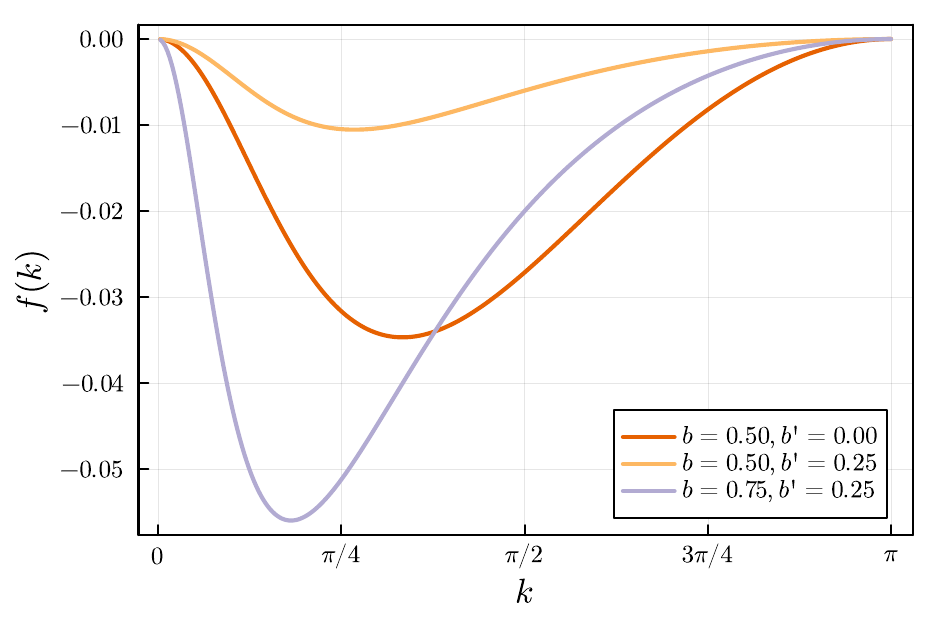}
    \caption{The function $f(k)=\ln(\cos(\varphi_k/2))$ for some typical values of $b$ and $b'$.}
    \label{fig_fk}
\end{figure}

Using the expression \eqref{eq_gs_b_and_bprim}, we find that
\begin{align}\label{eq_overlaps_sum}
    \ln \Big(\braket{\varphi_\pm(g)}{\varphi_\pm(g')}\Big )=\sum_{0\le k \le \pi} \ln \Big(\cos(\phi_k/2)\Big ),
\end{align}
where the difference between different parity sectors is in the values that $k$ takes. Let us define the function $f(k)$ for a fixed $g$ and $g'$  as follows,
\begin{align}\label{eq_f_def}
    f(k)\equiv \ln\Big(\cos(\phi_k/2)\Big).
\end{align}
The function $f(k)$ for few typical values of $g$ and $g'$ is plotted in Fig.\ref{fig_fk}. 
let us define $\Delta=2\pi/N$, to be the difference between the consecutive values of $k$ in \eqref{eq_k_values}. Accordingly, 
\begin{align}
    \frac{2\pi  \ln \Big(\braket{\varphi_\pm(g)}{\varphi_\pm(g')}\Big )}{N}&=\sum_{0\le k \le \pi} f(k)\Delta\\
    &=\int_0^\pi \dd k~f(k)+O(1/N).
\end{align}
This shows that the overlaps $\braket{\varphi_\pm(g)}{\varphi_\pm(g')}$ decay exponentially in $N$, with the rate given as the area under $f(k)$ (See Fig.\ref{fig_fk}). Note that $f(k)=0$ for $g'=g$. For $|g-g'|\ll 1$, one can expand $f(k)$ in terms of $(g-g')$ and would find that,
\begin{align}
    \braket{\varphi_\pm(g)}{\varphi_\pm(g')}=A_\pm\, \exp\Big( - \frac{(g-g')^2}{32(1-g^2)}  ~N \Big),
\end{align}
where $A_\pm=O(1)$. As we show in the next section $A_+=A_-$ up to corrections which vanish faster than any power in $1/N$.

\subsubsection{Computing the ratio $\frac{\braket{\varphi_+(g)}{\varphi_+(g')}}{\braket{\varphi_-(b)}{\varphi_-(g')}}$}
In this section we show that the ratio $\frac{\braket{\varphi_+(g)}{\varphi_+(g')}}{\braket{\varphi_-(b)}{\varphi_-(g')}}$ is equal to $1$ up to errors which vanish faster than any power in $1/N$. Based on the perturbation argument and backed by numerics, we expect the errors to vanish exponentially in $N$. 

Let $k_m=2\pi m/N$ denote the values of $k$ that the negative parity solution takes. Also remember $\Delta$ denotes $2\pi/N$ which is the difference between two consecutive values of $k_m$. Let $r(N)=\frac{\braket{\varphi_+(g)}{\varphi_+(g')}}{\braket{\varphi_-(g)}{\varphi_-(g')}}$ denote the ratio between the overlaps. Based on Eq. \eqref{eq_overlaps_sum} we may write $\ln(r)$ as,
\begin{align}
    \ln(r)&=\sum_{k_m}f(k_m+\frac{\Delta}{2})-\sum_{k_m}f(k_m)\nonumber \\
    &=\sum_{k_m}[f(k_m+\frac{\Delta}{2})-f(k_m)]\label{eq_lnr_exact_sum}\\
    &=\sum_{k_m} \sum_{j=1}^\infty \frac{f^{(j)}(k_m)}{j!}\frac{\Delta^j}{2^j}\nonumber\\
    &=\sum_{j=0}^\infty a_j \Delta^j \sum_{k_m} f^{(j+1)}(k_m)\Delta,
    \label{eq_ln_taylor_ex}
\end{align}
where function $f(k)$ is defined in Eq. \eqref{eq_f_def} and $f^{(j)}$ stands for its $j$'th derivative, and
\begin{align}
    a_j=\frac{1}{2^{j+1}~(j+1)!}.
\end{align}
Note that each $\sum_{k_m}$ in that expansion above converges to an integral for $N\to \infty$ and thus is of order $O(1)$, so one can think of the expansion in Eq. \eqref{eq_ln_taylor_ex} as an $1/N$ expansion of $\ln r$. To find the analytic form of each order, we may approximate the sum over $\sum_{k_m}$ with an integral $\int \dd k$. However, doing so would introduce subleading errors which we need to account for. In general, For a given function $h(x)$, let $H(x)=\int \dd x~ h(x)$, i.e. $H'(x)=h(x)$. Then, form the Taylor expansion of $H(x+\Delta)$ we have,
\begin{align}
    H(x+\Delta)-H(x)&=H'(x)\Delta + H''(x)\frac{\Delta^2}{2!}+\cdots\\
                    &=h(x)\Delta + h'(x) \frac{\Delta^2}{2!}+\cdots\\
                    &=h(x)\Delta +\sum_{j=1}^\infty \frac{h^{(j)}(x)}{(j+1)!}\Delta^{j+1}  
\end{align}
Therefore, assuming $\{x_i\}_{i=1}^M$ is a sequence of numbers with $x_{i+1}-x_i=\Delta$, we find,
\begin{align}
    \sum_{x_i} h(x_i)\Delta=&\int_{x_0}^{x_M+\Delta} \dd x~h(x) \nonumber\\
    &- \sum_{j=1}^\infty \frac{\Delta^{j}}{(j+1)!}\sum_{k_m} h^{(j)}(x_i)\Delta.
\end{align}
In particular, in our case we find that,
\begin{align}
    \sum_{k_m} f^{(n+1)}(k_m)\Delta=&\int_{0}^\pi \dd k~ f^{(n+1)}(k) \nonumber\\
    &- \sum_{j=1}^\infty b_{j}\Delta^j\sum_{k_m} f^{(n+j+1)}(k_m)\Delta,
\end{align}
with,
\begin{align}
    b_j=\frac{1}{(j+1)!}.
\end{align}
Now we can compute $\ln(r)$ to arbitrary order in $1/N$. We start by computing the zeroth order contribution,
\begin{align}
    \ln(r)&=a_0\sum_{k_m}f'(k_m)\Delta +O(\frac{1}{N})\\
        &=a_0\int_0^\pi \dd k~f'(k) +O(\frac{1}{N})\label{eq_replacing_sum_int_0}\\
        &=a_0\big[f(\pi)-f(0)\big] +O(\frac{1}{N})
\end{align}
To compute the $1/N$ contribution, in addition to $j=1$ term in Eq. \eqref{eq_ln_taylor_ex}, we have to take into account the $1/N$ correction that comes from replacing $a_0\sum_{k_m}f'(k_m)\Delta$ by $a_0 \int_0^\pi \dd k ~f(k)$ in Eq. \eqref{eq_replacing_sum_int_0}. Therefore we find that,
\begin{align}
    \ln(r)=&a_0\big[f(\pi)-f(0)\big]\nonumber\\
            &+(a_1-a_0\,b_1)\Delta\sum f''(k_m)\Delta +O(\frac{1}{N^2})\\
            =&a_0\big[f(\pi)-f(0)\big]\nonumber\\
            &+(a_1-a_0\,b_1)\Delta\int_0^\pi\dd k~ f''(k) +O(\frac{1}{N^2})\\
            =&a_0\big[f(\pi)-f(0)\big]\nonumber\\
            &+(a_1-a_0\,b_1)\Delta\big[f'(\pi)-f'(0)\big]+O(\frac{1}{N^2})
\end{align}
Following the same logic, one can see easily that the $1/N^2$ contribution would be $\Big(a_3-(a_1-a_0 b_1)b_1-a_0 b_2\Big)\Delta^2 \big[f''(\pi)-f''(0)\big] $, and in general if we keep the terms up to order $1/N^{M}$ we find,
\begin{align}\label{eq_lnr_N_exp}
    \ln(r)=\sum_{j=0}^{M} \alpha_j \Delta^j \big[f^{(j)}(\pi)-f^{(j)}(0)\big]+O\Big(\frac{1}{N^{M+1}}\Big),
\end{align}
with $\alpha_j$ defined via the following recursive relation:
\begin{align}
    &\alpha_j=a_j-\sum_{i=0}^{j-1} \alpha_i ~ b_{j-i},\label{eq_alpha_recersion}\\
    &\alpha_0=a_0.
\end{align}
The sum in Eq. \eqref{eq_alpha_recersion} comes from the integral approximation of the previous orders. Let us define $G_n$ such that
\begin{align}
    \alpha_j=\frac{G_{j+1}}{2^{j+1}~(j+1)!}.
\end{align}
After substituting values of $a_j$ and $b_j$ into Eq. \eqref{eq_alpha_recersion} and reordering the terms, we find that the $G_n$ numbers satisfy the following recurrence relation,
\begin{align}\label{eq_genocchi_rec}
    &\sum_{i=1}^j {j+1 \choose i} \frac{2^{j-i}}{j+1}~G_j=1,
\end{align}
which can be used to compute $G_n$ recursively for $n\ge 1$. The first few numbers in this sequence are,
\begin{align}
    1,-1,0,1,0,-3,0,17,0,-155,0,2073,\cdots
\end{align}
This number sequence is known as the Genocchi numbers of first kind\cite{A036968}. To show that the recurrence relation \eqref{eq_genocchi_rec} indeed corresponds to the Genocchi sequence, one can plug in $x=2$ and $k=1$ into recurrence relation (2.46) in Ref.\cite{horadam1991genocchi} and then use (2.21) and (2.22) to simplify it to get Eq. \eqref{eq_genocchi_rec}.
Importantly, $G_{2j+1}=0$ for all $j\ge 1$\cite{horadam1991genocchi}, which means that $\alpha_{2j}=0$ for all $j\ge 1$. Moreover, $f(\pi)=f(0)=0$, so the $j=0$ term in Eq. \eqref{eq_lnr_N_exp} is also zero. Lastly, it is straight forward to show that $f(k)=f(-k)$ and $f(\pi-k)=f(\pi+k)$, and hence $f^{(2j+1)}(0)=f^{(2j+1)}(\pi)=0$ for $j\ge 0$. Therefore, Eq. \eqref{eq_lnr_N_exp} shows that $\ln(r)$ is zero up to corrections of order $O(1/N^{M+1})$. Since $M$ is arbitrary, this shows that,
\begin{align}
    r(N)=\frac{\braket{\varphi_+(g)}{\varphi_+(g')}}{\braket{\varphi_-(g)}{\varphi_-(g')}} =1+\varepsilon(N),
\end{align}
where $\varepsilon(N)$ goes to zero faster than any power in $1/N$. Also note that the argument above works even if $|g-g'|$ scales with system size such that $|g-g'|\to 0$ in the thermodynamic limit. This in turn rules out scaling of the form $\ln(r)\sim \text{poly}(g,g')e^{-(g-g')^p N}$ for some $p>0$, because this scaling has a non-zero expansion in powers of $1/N$, for $|g-g'|\sim 1/N^p$. 

\begin{figure}
    \centering
    \includegraphics[width=\columnwidth]{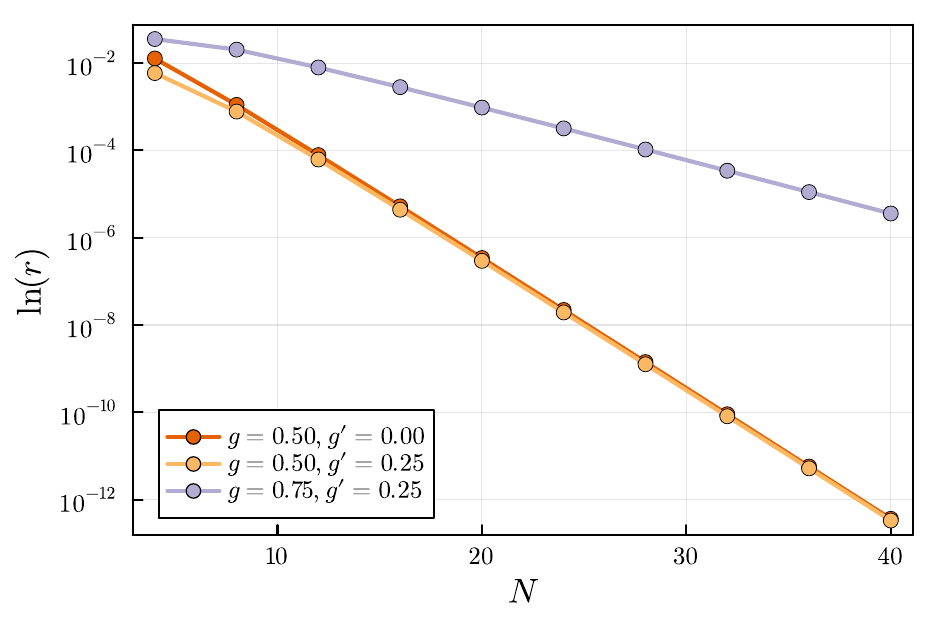}
    \caption{log of the ratio $\frac{\braket{\varphi_+(g)}{\varphi_+(g')}}{\braket{\varphi_-(g)}{\varphi_-(g')}}$ as a function of $N$ for some typical values of $g$ and $g'$.}
    \label{fig_lnr}
\end{figure}

If we compute $\ln(r)$ numerically using the exact expression in Eq. \eqref{eq_lnr_exact_sum}, we find that $\ln(r)$ decays exponentially with $N$ (see Fig.\ref{fig_lnr}). 

\section{Perturbation theory}\label{apx_perturb_theory_TFIM}
Consider the TFIC Hamiltonian,
\begin{align}
    H=-\sum_{i=1}^{N} Z_i Z_{i+1}-g \sum_{i=1}^{N} X_i=H_0+g~V.
\end{align}
Let $\ket{\varphi_+(g)}$ and $\ket{\varphi_-(g)}$ be ground states of the above Hamiltonian in the even and odd sectors, specified by the parity operator $P=\prod_{i=1}^N X_i$. Moreover, assume we fix the phases of $\ket{\varphi_\pm(g)}$ via requiring $\braket{\varphi_\pm(g)}{\partial_g \varphi_\pm (g)}=0$. In this section, we are going to use perturbation theory to show, among others, $ \mel{\ghz_-}{e_\bmm}{\varphi_-(g)}\simeq \mel{\ghz_+}{e_\bmm}{\varphi_+(g)}$ up to corrections which are of order $\mathcal{O}(g^{N/2})$. 

We are going to treat the the transverse field term, $g\, V=g\sum_i X_i=g\,V$, as a perturbation to the fixed point Hamiltonian $H_0$. Before going into details, it is worth commenting on the validity of perturbative results. Note that $\norm{gV}=O(g\, N)$, so naively one might expect the perturbation theory to be valid only if $g\ll 1/N$,i.e. a neighborhood whcih vanishes in the thermodynamic limit. This would be indeed true if one used first order (or any finite order) pertubation theory. Consder the perturbative expansion of the ground state energy as an example,
\begin{align}
    E_0=E_0^{(0)}+g\,E_0^{(1)}+g^2\, E_0^{(2)}+\cdots
\end{align}
In general, each $E_0^{(n)}$ coefficient is a function of $N$, and could diverge as one takes the thermodynamic limit. Therefore, if one keeps the terms in the expansion up to some finite order (independent of $N$),  the error in the expansion would not be controlled. However, if one keeps all the terms up to a macroscopic order, say $N$, then the error of cutting off the expansion would be $~c_N\, g^{N+1}$, for some coefficient $c_N$ which is a function of $N$. Now, if $c_N$ grows at most exponentially with $N$, there exists a finite neighborhood around $g=0$ where such a perturbative expansion can be trusted. In what follows, we will look at the $n$-th order perturbation theory with $n=O(N)$. We do not prove that error coefficients will not diverge super-exponentially in $N$, but we use numerical simulations to justify the perturbative expansion. While this treatment cannot be considered rigorous, we believe it explains the essential physics of the problem.   

Since $[V,P]=0$, at any order of perturbation theory the odd and even parity sectors remain detached from one another. As such, we may focus on each sector separately and use non-degenerate perturbation theory to find the corrections to the unperturbed ground state on that sector. 
Let us focus on $P=+1$, and to simplify the notation let us denote the corresponding ground state simply by $\ket{\varphi}\equiv\ket{\varphi_+(g)}$. We also use $\ket{\bmm}$ to denote $e_\bmm\ket{\ghz_+}$. Note that $\{\ket{\bmm}~|~ \bmm \in \mathbb{Z}_2^{N-1}\}$ is a complete basis set for the $P=+1$ subspace of the Hilbert space. They are also eigenstates of $H_0$. 
We use $E^{(0)}_\bmm$ to denote the energy of $\ket{\bmm}$ with respect to $H_0$, i.e. 
\begin{align}
    H_0\ket{\bmm}=E^{(0)}_\bmm \ket{\bmm}.
\end{align}
Consider the following perturbative expansion for $\ket{\varphi}$:
\begin{align}
\ket{\varphi}=\ket{\varphi^{(0)}}+g\ket{\varphi^{(1)}}+g^2\ket{\varphi^{(2)}}+\cdots.
\end{align}
Note that $\ket{\varphi^{(0)}}=\ket{\ghz_+}=\ket{\bm{0}}$. We have a similar expression for the ground state energy:
\begin{align}
    E_0=E_0^{(0)}+g\,E_0^{(1)}+g^2\, E_0^{(2)}+\cdots
\end{align}
and we have, 
\begin{align}\label{eq_sh_eq}
    (H_0+g\,V)\ket{\varphi}=E_0\ket{\varphi}.
\end{align}
The $n$'th order contributions to Eq. \eqref{eq_sh_eq} then yields, 
\begin{align}
    &H_0 \ket{\varphi^{(n)}}+V\ket{\varphi^{(n-1)}}\nonumber\\
    &=E^{(0)}_0\ket{\varphi^{(n)}}+E^{(1)}_0\ket{\varphi^{(n-1)}}+\cdots+E^{(n)}_0\ket{\varphi^{(0)}},
\end{align}
which can be rewritten as,
\begin{align}\label{eq_nth_order_she}
    &\qty(H_0-E_0^{(0)})\ket{\varphi^{(n)}}=\nonumber \\
    &-V\ket{\varphi^{(n-1)}}+E_0^{(2)}\ket{\varphi^{(n-2)}}+\cdots+E^{(n)}_0\ket{\varphi^{(0)}},
\end{align}
where we have used $E_0^{(1)}=\mel{\varphi^{(0)}}{V}{\varphi^{(0)}}=0$. 
By multiplying both sides with $\bra{\varphi^{(0)}}$, we find, 
\begin{align}\label{eq_E0n_recursive}
    E_0^{(n)}=&\mel{\varphi^{(0)}}{V}{\varphi^{(n-1)}}-E_0^{(2)}\braket{\varphi^{(0)}}{\varphi^{(n-2)}}\nonumber\\
    &-E_0^{(3)}\braket{\varphi^{(0)}}{\varphi^{(n-3)}}\cdots-E_0^{(n-1)}\braket{\varphi^{(0)}}{\varphi^{(1)}}.
\end{align}
As for finding $\ket{\varphi^{(n)}}$, we may expand it in terms of $\ket{\bmm}$ states,
\begin{align}
    \ket{\varphi^{(n)}}=\sum_\bmm c_\bmm^{(n)}\ket{\bmm},
\end{align}
and for $\bmm \ne \bm{0}$, can find $c_\bmm^{(n)}$ by multiplying both sides of  Eq. \eqref{eq_nth_order_she} by $\bra{\bmm}$
\begin{align}\label{eq_cnm_recursive}
    c_\bmm^{(n)}=&\frac{1}{E^{(0)}_\bmm-E_0^{(0)}}\Bigg[-\mel{\bmm}{V}{\varphi^{(n-1)}}\nonumber\\
    &+E_0^{(2)}\braket{\bmm}{\varphi^{(n-2)}}+\cdots E_0^{(n)}\braket{\bmm}{\varphi^{(0)}}
    \Bigg].
\end{align}
To find $c_{\bm{0}}^{(n)}=\braket{\varphi^{(0)}}{\varphi^{(n)}}$ we require that $\braket{\varphi}{\partial_g \varphi}=0$ at each order of perturbation theory. It then yields,
\begin{align}\label{eq_cn0_recursive}
    n\braket{\varphi^{(0)}}{\varphi^{(n)}}&+(n-1)\braket{\varphi^{(1)}}{\varphi^{(n-1)}}+\cdots\nonumber\\
    &+2\braket{\varphi^{(n-2)}}{\varphi^{(2)}}+\braket{\varphi^{(n-1)}}{\varphi^{(1)}}=0.
\end{align}
Note that it ensures $\ket{\varphi}$ to be normalized as well. 

\begin{figure}
    \centering
    \includegraphics[width=\columnwidth]{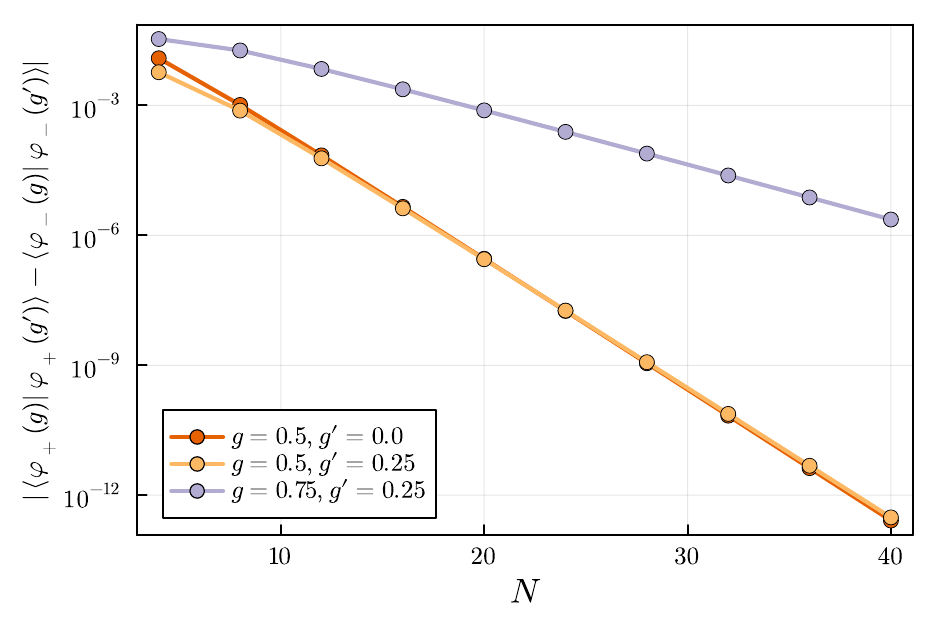}
    \caption{The difference $|\braket{\varphi_+(g)}{\varphi_+(g')}-\braket{\varphi_-(g)}{\varphi_-(g')}|$ as a function of $N$ for few different choices of $g$ and $g'$.}
    \label{fig_overlap_delta}
\end{figure}

Eq. \eqref{eq_cnm_recursive} along side Eq. \eqref{eq_cn0_recursive} and Eq. \eqref{eq_E0n_recursive} can be used to find $n$'th order corrections to the ground state wave function based on the corrections up to the $n-1$'th order. By rearranging the terms we may write the ground state as,
\begin{align}
    \ket{\varphi}=\sum_{\bmm \in \mathbb{Z}_2^{N-1}}c_\bmm \ket{\bmm},
\end{align}
with,
\begin{align}
    c_\bmm=\sum_{n=0}^\infty c_\bmm^{(n)} g^n.
\end{align}
In particular, note that at zeroth order, only $\ket{\bm{0}}=\ket{\ghz_+}$ appears in the expansion of $\ket{\varphi}$. At first order, only $\ket{\bm{0}}$ and $\ket{\bmm}$ states with $e_\bmm =X_j$ for some $j$ appear in the expansion. At second order, only $\ket{\bmm}$ with $|\text{Supp}(e_\bmm)|\le 2$ appear in the expansion and so on. 
More generally, one can easily use induction to show that at $n$'th order,
\begin{align}\label{eq_cmn_is_zero_k_g_n}
    c^{(n)}_\bmm=0\qquad\text{for}\qquad |\text{Supp}(e_\bmm)|>n.
\end{align}
Moreover, while we considered the $P=+1$ sector so far, the exact same holds for $P=-1$ sector; the only modification is that $\ket{\varphi}$ denotes $\ket{\varphi_-(g)}$ and $\ket{\bmm}$ denotes $e_\bmm\ket{\ghz_-}$ instead. Importantly, the $c_\bmm^{(n)}$ coefficients will be the same for the $P=+1$ and $P=-1$ sectors up to order $N/2-1$. This is simply because up to that order, the recursive equations determining $c_\bmm^{(n)}$ are exactly the same in the odd and even parity sectors. It is only at order $n\ge N/2$, when the $\mel{\bmm}{V}{\varphi^{(n-1)}}$ term in Eq. \eqref{eq_cnm_recursive} becomes different for the two sectors and hence $c_\bmm^{(n)}$ starts to be different in the two sectors. Since $c_\bmm=\braket{\bmm}{\varphi}=\mel{\ghz}{e_\bmm}{\varphi}$, we find that,
\begin{align}\label{eq_mel_gpm_em}
    \mel{\ghz_-}{e_\bmm}{\varphi_-(g)}=\mel{\ghz_+}{e_\bmm}{\varphi(g)_+}+\mathcal{O}(g^{N/2}).
\end{align}
Eq. \eqref{eq_mel_gpm_em} by itself is not particularly useful given that $\mel{\ghz_\pm}{e_\bmm}{\varphi_\pm(g)}$ are exponentially small themselves. However, one could argue that they are parametrically larger than the error term in Eq. \eqref{eq_mel_gpm_em}. Let $k=|\text{Supp}(e_\bmm)|$, and assume $k/N=a$ for $0<a<\frac{1}{2}$. It is straightforward to  use \eqref{eq_cmn_is_zero_k_g_n} and Eq. \eqref{eq_cnm_recursive} to show that, for both $P=\pm 1$ sectors, $c^{(k')}_\bmm=0$ for $k'<k$, but $c^{(k)}_\bmm\ne 0$. This in turn suggests that 
$\mel{\ghz_\pm}{e_\bmm}{\varphi_\pm(g)}\sim g^{k}$, and hence, 
\begin{align}
    &\mel{\ghz_-}{e_\bmm}{\varphi_-(g)}=\mel{\ghz_+}{e_\bmm}{\varphi_+(g)}(1+\mathcal{O}(g^{N/2-k})).
\end{align}
\begin{figure}
    \centering
    \includegraphics[width=\columnwidth]{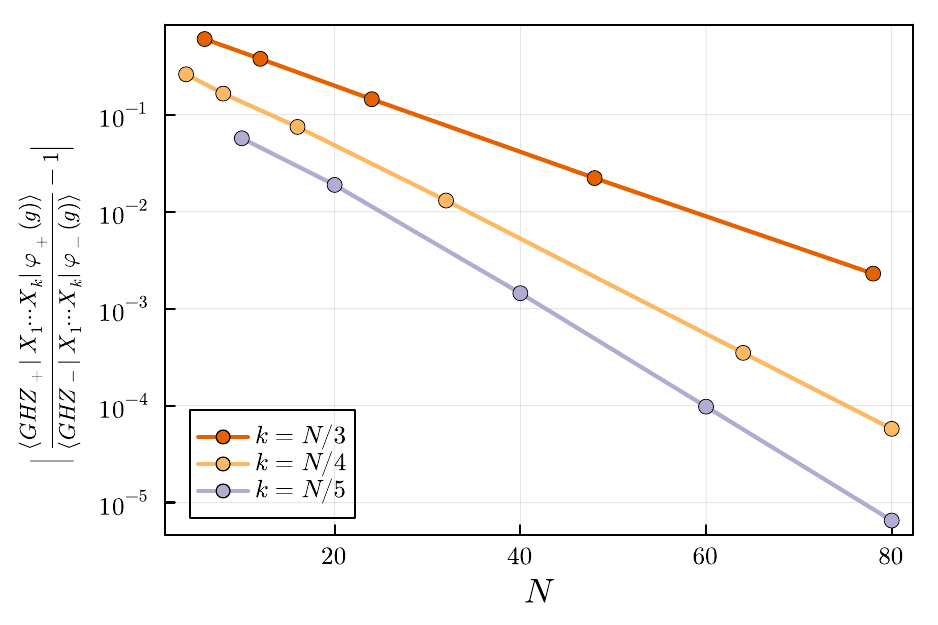}
    \caption{The difference $|\frac{\mel{\ghz_+}{e_\bmm}{\varphi_+(g)}}{\mel{\ghz_-}{e_\bmm}{\varphi_-(g)}}-1|$ as a function of $N$, when $e_\bmm$ is taken to be $\Pi_{j=1}^k X_j$, for few different choices of $k$.}
    \label{fig_em_ratio}
\end{figure}
Fig.\ref{fig_em_ratio} shows the absolute value of $\frac{\mel{\ghz_+}{e_\bmm}{\varphi_+(g)}}{\mel{\ghz_-}{e_\bmm}{\varphi_-(g)}}-1$ as a function of $N$, when $e_\bmm=\Pi_{j=1}^k X_j$ for few different choices of $k=O(N)$. As is clear from the figure, the difference goes to zero exponentially fast in system size.
A similar line of argument shows that 
\begin{align}
    \braket{\varphi_+(g)}{\varphi_+(g')}=\braket{\varphi_-(g)}{\varphi_-(g')}+\mathcal{O}(g_\text{max}^{N}),
\end{align}
where $g_\text{max}=\max(g,g')$. To see this, note that the first term in the perturbative expansion of $\ket{\varphi_+(g)}$ that is has a different coefficient from the corresponding term in $\ket{\varphi_-(g)}$, is $c_\bmm^{(N/2)}$ with $k=|\text{Supp}(e_\bmm)|=N/2$. Therefore the first term in the perturbative expansion of $\braket{\varphi_+(g)}{\varphi_+(g')}$ which has a different value from the corresponding term in the expansion of $\braket{\varphi_-(g)}{\varphi_-(g')}$ is of order $(g'g)^{N/2}$. At the next order, one can see that $c_\bmm^{(N/2+1)}$ would be different for odd and even parity sectors, even for $k=|\text{Supp}(e_\bmm)|=N/2-1$. This then result in a terms of order $g'^{N/2-1}g^{N/2+1}$ and $g'^{N/2+1}g^{N/2-1}$ in the expansion of $\braket{\varphi_+(g)}{\varphi_+(g')}-\braket{\varphi_-(g)}{\varphi_-(g')}$. Continuing the same line of argument up to $N$th order shows that the terms that appear in the perturbative expansion of $\braket{\varphi_+(g)}{\varphi_+(g')}-\braket{\varphi_-(g)}{\varphi_-(g')}$, are of the order $g'^{N/2-m} g^{N/2+m}$ for $m=0,\cdots, N/2$. For fixed $g$,$g'$ and large $N$, one can see that the largest contribution comes from $g_\text{max}^N$. Fig.\ref{fig_overlap_delta} shows $|\braket{\varphi_+(g)}{\varphi_+(g')}-\braket{\varphi_-(g)}{\varphi_-(g')}|$ computed numerically for different choices of $g$ and $g'$, which clearly shows the exponential drop with $N$.
\section{Adiabatic Drift and Recovery for the Non-Integrable TFIC}\label{apx_nonintg_tfim}

Here we consider the problem of recoverability in the phase diagram of the following Hamiltonian,
\begin{align}\label{eq_nonintgrable_tfim}
    H(g)=&-\sum_i Z_i Z_{i+1} -\sum_i Z_i Z_{i+2}\nonumber\\
    &-g \sum_i X_i -g \sum_i X_i X_{i+1}.
\end{align}
This model no longer maps to free fermions under JW transformation. Nonetheless,  the $X \longleftrightarrow ZZ$ duality still pins the critical point at $b=1$, separating the ferromagnetic phase   ($g<1$) from the paramagnetic phase ($g>1$). Moreover, the ground space subspace at the fixed point $g=0$ still coincides with the the codespace of the repetition code. We use this model to show that the perfect recoverability in Section \ref{sec_repetition_code_and_1dTFIM} does not depend on the integrability of the underlying Hamiltonian. 

\begin{figure}
    \centering
    \includegraphics[width=\columnwidth]{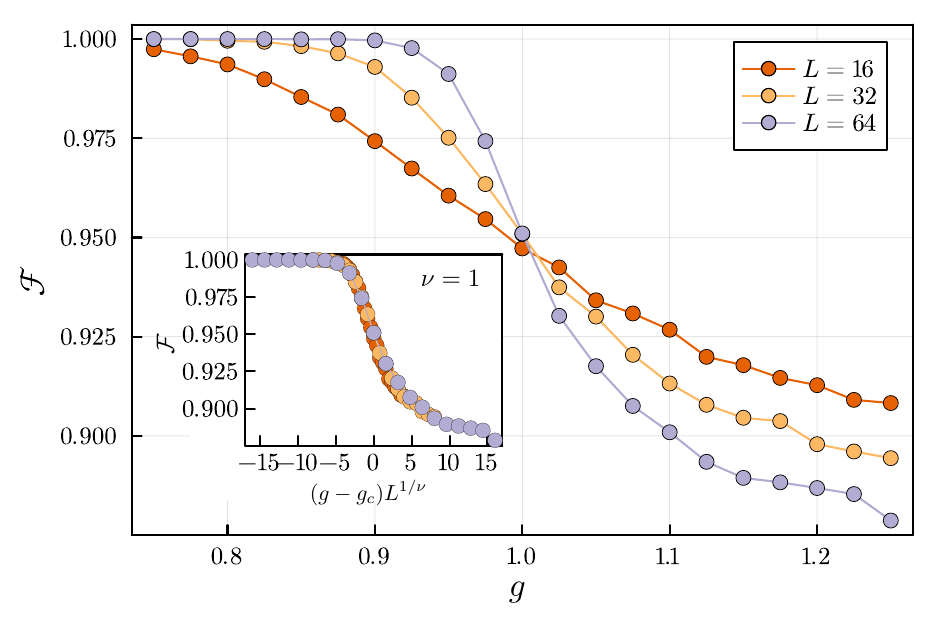}
    \caption{Fidelity of the recovered state with respect to the original encoded state as a function of $g$, for the nonintegrable model described by the Hamiltonian \eqref{eq_nonintgrable_tfim}. The inset shows the data collapse of the same data points, using $g_c=1$ and $\nu=1$.}
    \label{fig_isingFidelity_nonintg}
\end{figure}

We use matrix product state (MPS) representation to numerically find the odd and even parity ground states $\ket{\varphi_\pm(g)}$ of Hamitonian \eqref{eq_nonintgrable_tfim} on an open chain of $L$ qubits. To fix the relative phase of $\ket{\varphi_\pm(g)}$ according to the adiabatic evolution, we fix $\braket{\ghz_\pm}{\varphi_\pm(g)}$ to be real, which according to Eq. \eqref{eq_mel_gpm_em} only introduces an error which is exponentially small in $L$. We uses the same initial code state $\ket{\psi}=1/\sqrt{2}(\ket{\ghz_+}+\ket{\ghz_-})$, and we use the same recovery map after the adiabatic evolution as in Section \ref{sec_repetition_code_and_1dTFIM}. As can be seen from Fig.~\ref{fig_isingFidelity_nonintg} the recovery succeeds perfectly in the thermodynamic limit. The inset shows the data collapse with $g_c=1$ and $\nu=1$, which are the same as for the integrable case. Nonetheless, note that the fidelity at the critical point is lower from its critical value in Fig.~\ref{fig_isingFidelity_c}.

\section{Computing $\calZ_i(\beta)$}\label{apx_calZ_derivation}
In this Section we compute $\calZ_i(\beta)$ appearing in Section \ref{sec_toric_code}. For the $\ket{\varphi_i(\beta)}$ states to be normalized we have,
\begin{align}\label{eq_C4_partition_wf_expval}
    \calZ_i(\beta)=\expval{\exp(\beta \sum_j Z_j)}{\varphi_i(0)}.
\end{align}
We first compute the expectation value for $\ket{\varphi_0(0)}$. First note that one may write this state as follows,
\begin{align}
    \ket{\varphi_0(0)}&=\frac{1}{2^{\frac{M}{2}}}\prod_s(1+A_s)\ket{0}\\
    &=\frac{1}{2^{\frac{M}{2}}}\sum_{\sigma\in\{-1,1\}^M} \prod_s A_s^{(1-\sigma_s)/2}\ket{0},
\end{align}
where $\ket{0}=\ket{0}^{\otimes N}$ is the state where all qubits are in the $\ket{0}$ of Pauli-$Z$, and $M$ is the number of stars  in $\Lambda$. We can think of $\sigma_s$ as a classical spin sitting on the star $s$. If the qubit $j$ neighbors the stars $\mu$ and $\nu$ as in Fig.~\ref{fig:2dising_toric}, then,
\begin{align}
    Z_j \prod_s A_s^{(1-\sigma_s)/2}\ket{0}=\sigma_{\mu} \sigma_{\nu} \prod_s A_s^{(1-\sigma_s)/2}\ket{0},
\end{align}
Therefore, any configuration of classical spins $\{\sigma_s\}$ corresponds to a specific computational basis of qubits, with the state of the qubit given by the product of the neighboring classical spins. Fixing the boundary classical spins to be in the $\sigma=1$ state ensures this to be true for boundary qubits as well. Also note that the states represented by different classical spins are orthogonal. Therefor we find,
\begin{align}
    \calZ_0(\beta)=\frac{1}{2^M}\sum_{\sigma\in\{-1,1\}^M}\exp(\beta \sum_{\langle \mu,\nu \rangle} \sigma_\mu \sigma_\nu),
\end{align}
which is simply the partition function of the 2D-classical Ising model.

As for $\calZ_1(\beta)$, one can easily verify that the same calculation works for this case as well, if one replaces $\ket{0}$ by $\overline{X}\ket{0}$, where all qubits are in the $\ket{0}$ state, except the ones on the support of $\overline{X}$, which are flipped to $\ket{1}$. This then implies that for the state  $\prod_s A_s^{(1-\sigma_s)/2}\overline{X}\ket{0}$, we have $Z_j=\sigma_{\mu}\sigma_{\nu}$ everywhere except the qubits on the support of $\overline{X}$, for which we have $Z_j=-\sigma_{\mu}\sigma_{\nu}$. Thus $\calZ_1(\beta)$ would be the partition function of the 2D-classical Ising model with a domain wall inserted along the support of  $\overline{X}$ (see Fig.~\ref{fig:ising_domainwall}), meaning that the coupling along the support of $\overline{X}$ is flipped to be anti-ferromagnetic.

\section{Explicit summations of Eq. \eqref{eq_N_0_sum} and Eq. \eqref{eq_N_1_sum}}\label{apx_explicit_summation}
In this section, we explain how one can arrive at Eq. \eqref{eq_N_0_sum} and Eq. \eqref{eq_N_1_sum} by exlicitly performing the sums on the left hand sides. 

\subsection{relation between $P[e]$ and the partition function of the random-bond Ising model(RBIM)}
First we describe the relation between the function $P([e])$ and the partition function of the random-bond Ising model, as was noted in Ref.\cite{dennis2002topological}. define $J$ as
\begin{align}\label{eq_nishimori}
    e^{-2J}=\frac{q}{1-q},
\end{align}
so we have,
\begin{align}
    P[e]=(1-q)^N\sum_{e'\in [e]} e^{-2J|e'|}.
\end{align}
Note that Eq. \eqref{eq_nishimori} is just the equality which defines the Nishimori line\cite{nishimori1981internal}. For completeness we repeat Eq. \eqref{eq_q_vs_beta} here as well, which gives $q$ in terms of $\beta$,
\begin{align}\label{eq_q_vs_beta_apx}
    q=\frac{1-e^{-\beta}}{2},
\end{align}

If $e'\in[e]$, it means that $C=e+e'$ represents a $Z$-stabilizer and hence $Z(C)$ can be written as the product of plaquette operators. If we introduce a spin $\tau_p=\pm 1$ on each plaquette, we can represent any product of plaquette stabilizers by a spin configuration $\{\tau_p\}$, such that $Z(C)=\prod_p B_p^{(1-\tau_p)/2}$. It is easy to see that the corresponding $C$ has the following elements,
\begin{align}
    C_j=(1-\tau_\mu \tau_\nu)/2,
\end{align}
where $\mu$ and $\nu$ are the plaquettes that are neighboring the edge $j$. Since the product of all plaquettes is identity, spin configurations $\{\tau_p\}$ and $\{-\tau_p\}$ correspond to the same loop $C$. Given $C$, we can  construct $e'$ as $e'_j=e_j+C_j$, however the sum here is modulo $2$. To take care of that, we introduce variable $\eta_{j}=(1-2e_j)=\pm 1$. It is more convenient to write $\nu_j$ as $\eta_{\mu \nu}$ where $\mu$ and $\nu$ are the plaquettes neighboring the edge $j$. Then we have,
\begin{align}
    e'_j&=\big[1-(1-2e_j)(1-2C_j)\big]/2\nonumber\\
    &=(1-\eta_{\mu\nu}\tau_\mu \tau_\nu)/2
\end{align}
where now all the summations are ordinary sums. Accordingly we find,
\begin{align}
    P([e])&=\frac{(1-q)^N}{2}\sum_{\tau\in\{1,-1\}^{M'}}e^{-J\sum_{\langle \mu,\nu \rangle}(1-\eta_{\mu \nu}\tau_\mu \tau_\nu)}\nonumber\\
    &=\frac{[q(1-q)]^{N/2}}{2}\sum_{\tau\in\{1,-1\}^{M'}}e^{J\sum_{\langle \mu,\nu \rangle}\eta_{\mu \nu}\tau_\mu \tau_\nu}\nonumber\\
    &=\frac{[q(1-q)]^{N/2}}{2} \calZ_\text{RBIM}(e,J).
\end{align}
where the factor of $\frac{1}{2}$ accounts for the double redundancy in representation of $e'$ in terms of $\tau$ spins. The $e$ error determines the disorder $\eta$ in $\calZ_\RBIM$. Note that if $e'\in [e]$ then $\calZ_\RBIM(e,J)=\calZ_\RBIM(e',J)$. 
\subsection{Relation between $\overline{\calZ_\RBIM(e,J)}$ and $\calZ_\text{Ising}$}
Here we related the average of partition function of random bond Ising model the partition function of Ising model. It goes as this,
\begin{align}
    &\overline{\calZ_\RBIM(e,J)}=\overline{\sum_{\tau\in\{1,-1\}^{M'}}e^{J\sum_{\langle \mu,\nu \rangle}\eta_{\mu \nu}\tau_\mu \tau_\nu}}\\
    &=\sum_{\tau\in\{1,-1\}^{M'}}\prod_{{\langle \mu,\nu \rangle}}\overline{\cosh(J)+\sinh(J)\eta_{\mu \nu}\tau_\mu \tau_\nu}\\
    &=\sum_{\tau\in\{1,-1\}^{M'}}\prod_{{\langle \mu,\nu \rangle}}\cosh(J)+\sinh(J)(1-2q)\tau_\mu \tau_\nu\\
    &=\Bigg[\frac{\cosh(J)}{\cosh(\tilde{J})}\Bigg]^{N}\sum_{\tau\in\{1,-1\}^{M'}}e^{\tilde{J}\sum_{\langle \mu,\nu \rangle}\tau_\mu \tau_\nu}\\
    &=\Bigg[\frac{\cosh(J)}{\cosh(\tilde{J})}\Bigg]^{N} \calZ_\Ising(\tilde{J})\label{eq_Zising_vs_Zrbim}.
\end{align}
with $\tilde{J}$ defined as,
\begin{align}\label{eq_Jtilde_vs_J}
    \tanh(\tilde{J})=(1-2q)\tanh(J).
\end{align}
It is also straightforward to see that 
\begin{align}
    \overline{\calZ_\RBIM(e+\ell_1,J)}=\Bigg[\frac{\cosh(J)}{\cosh(\tilde{J})}\Bigg]^{N} \calZ_{\Ising,\ell_1}(\tilde{J}).
\end{align}
where $ \calZ_{\Ising,\ell_1}$ is the Ising model with a domain wall along $\ell_1$, meaning that the coupling along $\ell_1$ are anti-ferromagnetic. This is simply because $\calZ_\RBIM(e+\ell_1,J)$ can be thought of as $\calZ_\RBIM(e,J)$ but where interactions along the links on $\ell_1$ are $-\eta_{\mu \nu} \tau_\mu \tau_\nu$ instead of $+\eta_{\mu \nu} \tau_\mu \tau_\nu$.

\subsection{The relation between $\calZ_\Ising$ and $\calZ_0$ and $\calZ_1$}

\begin{figure}
 \begin{subfigure}{0.49\columnwidth}
     \includegraphics[width=0.8\textwidth]{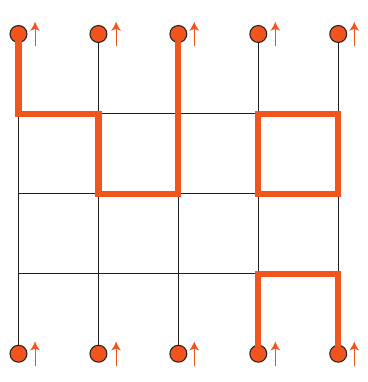}
     \caption{}
     \label{fig:hte_even}
 \end{subfigure}
 \hfill
 \begin{subfigure}{0.49\columnwidth}
     \includegraphics[width=0.8\textwidth]{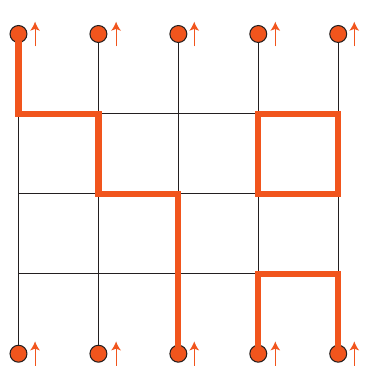}
     \caption{}
     \label{fig:hte_odd}
 \end{subfigure}
 \caption{(a) a loop configuration $c \in \mathcal{L}_\text{even}$ and (b) the corresponding loop configuration $c'=c+\ell_1\in \mathcal{L}_\text{odd}$ where $\ell_1$ is taken to be the line that runs vertically in the middle.}
 \label{fig_hte}
\end{figure}

Note that $\calZ_\Ising$ is the partition function for $\tau$ spins which has free open boundary conditions on top and bottom of cylinder. But $\calZ_0$ and $\calZ_1$ are partition function for $\sigma$ spins, with fixed boundary conditions on top and bottom. As we will see these partition functions are related to each other via the high-temperature low-temperature duality of the Ising model. We start by high-temperature expansion of $\calZ_0$,
\begin{align}
    &\calZ_0(\beta)=\frac{1}{2^M}\sum_{\sigma\in\{-1,1\}^M}\exp(\beta \sum_{\langle \mu,\nu \rangle} \sigma_\mu \sigma_\nu)\\
    &=\frac{1}{2^M}\sum_{\sigma\in\{-1,1\}^M}\prod_{\langle \mu,\nu \rangle} \Bigg[\cosh(\beta)+\sinh(\beta) \sigma_\mu \sigma_\nu \Bigg]\\
    &=\frac{1}{2^M}\cosh(\beta)^N 2^M \sum_{c\in \mathcal{L}}\tanh(\beta)^{|c|}
\end{align}
where $\mathcal{L}$ is the set of loop configurations with no open end inside the bulk, but they can end on the boundaries (see Fig.~\ref{fig_hte}). This is because we are not summing over the spins at the boundary so those terms are also allowed in addition to loop configurations. We can partition $\mathcal{L}$ into $\mathcal{L}=\mathcal{L}_\text{even}\cup \mathcal{L}_\text{odd}$ where $\mathcal{L}_\text{even}$ ($\mathcal{L}_\text{odd}$) is the set of allowed loop configurations with even (odd) lines at each boundary (see Fig.~\ref{fig:hte_even} and Fig.~\ref{fig:hte_odd}). Therefore we have,
\begin{align}
    &\calZ_0(\beta)=\cosh(\beta)^N\Bigg[\sum_{c\in \mathcal{L}_\text{even}}\tanh(\beta)^{|c|}+\sum_{c\in \mathcal{L}_\text{odd}}\tanh(\beta)^{|c|}\Bigg]\label{eq_ising_loop_expansion}\\
    &=\cosh(\beta)^N\Bigg[\sum_{c\in \mathcal{L}_\text{even}}\tanh(\beta)^{|c|}+\sum_{c\in \mathcal{L}_\text{even}}\tanh(\beta)^{|c+\ell_1|}\Bigg]
\end{align}

On the other hand, if one consider the partition function $\calZ_\Ising(\tilde{\beta})$ of the Ising model on $\tau$ spins that live on the plaquettes, with free boundary conditions on the top and bottom, the domain wall of each $\{\tau\}$ spin configuration correspond to a loop configuration $c \in \mathcal{L}_\text{even}$, whose energy $\tilde{\beta}\sum_{\langle \mu,\nu \rangle}\tau_\mu \tau_\nu=\tilde{\beta}\, N-2\tilde{\beta}|c|$.Therefore, 
\begin{align}
    \calZ_\Ising(\tilde{\beta})&=\sum_{\tau\in\{1,-1\}^{M'}}e^{\tilde{J}\sum_{\langle \mu,\nu \rangle}\tau_\mu \tau_\nu}\\
    &=2e^{\tilde{\beta}\,N}\sum_{c\in\mathcal{L}_\text{even}}\Bigg(e^{-2\tilde{\beta}}\Bigg)^{|c|},
\end{align}
where the factor of $2$ is there to account for the fact that each loop configuration corresponds to two spin configurations. One can similarly observe that $\calZ_{\Ising,\ell_1}(\tilde{\beta})$ can be written as,
\begin{align}
    \calZ_{\Ising,\ell_1}(\tilde{\beta})&=2e^{\tilde{\beta}\,N}\sum_{c\in\mathcal{L}_\text{even}}\Bigg(e^{-2\tilde{\beta}}\Bigg)^{|c+\ell_1|}.
\end{align}
Therefore, we see that if we define $\tilde{\beta}$ such that, 
\begin{align}\label{eq_betatilde_vs_beta}
    e^{-2\tilde{\beta}}=\tanh{\beta},
\end{align}
then we have, 
\begin{align}\label{eq_Z0_vs_Zising}
    \calZ_0=\frac{\cosh(\beta)^N}{2 e^{N\tilde{\beta}}}\Bigg[ \calZ_{\Ising}(\tilde{\beta})+\calZ_{\Ising,\ell_1}(\tilde{\beta})\Bigg]
\end{align}
As for $\mathcal{Z}_1$, note that one can move the domain wall around by flipping the $\sigma$ spins without changing $\mathcal{Z}_1$. Therefore, one can get rid of the domain wall completely at the expense of flipping all the fixed spins at the bottom boundary, such that they get fixed at $\sigma=-1$. Then, one can do the same loop expansion as in Eq. \eqref{eq_ising_loop_expansion}, but with the difference that the loops $c\in \mathcal{L}_\text{odd}$ appear with a minus sign. therefore, we find that, 
\begin{align}\label{eq_Z1_vs_Zising}
    \calZ_1=\frac{\cosh(\beta)^N}{2 e^{N\tilde{\beta}}}\Bigg[ \calZ_{\Ising}(\tilde{\beta})-\calZ_{\Ising,\ell_1}(\tilde{\beta})\Bigg]
\end{align}
\subsection{Deriving Eq. \eqref{eq_N_0_sum} and Eq. \eqref{eq_N_1_sum}}
The sums in Eq. \eqref{eq_N_0_sum} and Eq. \eqref{eq_N_1_sum} have two parts,
\begin{align}
    &\sum_\bmm (P([e_\bmm])+(-1)^j P([e_\bmm+\ell_1]))^2\nonumber\\
    &=\sum_{\bmm,k} P([e_\bmm+\ell_k])^2 +(-1)^j 2\sum_{\bmm}P([e_\bmm])P([e_\bmm+\ell_1]). 
\end{align}
As for the first part we have,
\begin{align}
    &\sum_{\bmm,k} P([e_\bmm+\ell_k])^2\nonumber\\
    &=\sum_{\bmm,k}\sum_{e'\in[e_\bmm + \ell_k]} (1-q)^{N-|e'|}q^{|e'|} P([e_\bmm+\ell_k])\nonumber\\
    &=\sum_{\bmm,k}\sum_{e'\in[e_\bmm + \ell_k]} (1-q)^{N-|e'|}q^{|e'|}P([e'])\\
    &=\sum_{e'} (1-q)^{N-|e'|}q^{|e'|}P([e'])=\overline{P([e])}\nonumber\\
    &=\frac{[q(1-q)]^{N/2}}{2}\overline{\calZ_\RBIM(e,J)}.
\end{align}

As for the second sum, we find that,
\begin{align}
    &\sum_{\bmm} P([e_\bmm])P([e_\bmm+\ell_1])\nonumber\\
    &=\frac{1}{2}\sum_{\bmm,k}P([e_\bmm+\ell_k])P([e_\bmm+\ell_k+\ell_1])\nonumber\\
    &=\frac{1}{2}\sum_{\bmm,k}\sum_{e'\in[e_\bmm + \ell_k]} (1-q)^{N-|e'|}q^{|e'|}P([e'+\ell_1])\\
    &=\frac{1}{2}\sum_{e'} (1-q)^{N-|e'|}q^{|e'|}P([e'+\ell_1])=\frac{1}{2}\overline{P([e+\ell_1])}\nonumber\\
    &=\frac{[q(1-q)]^{N/2}}{4}\overline{\calZ_\RBIM(e+\ell_1,J)}.
\end{align}
We note that $\calZ_\RBIM(e+\ell_1,J)$ can be equivalently thought of as the random bond Ising model partition function when there is a domain wall inserted along the support of $\overline{Z}$. So we find that, 
\begin{align}
    &\sum_\bmm (P([e_\bmm])+(-1)^j P([e_\bmm+\ell_1]))^2\nonumber\\
    &=\frac{[q(1-q)]^{N/2}}{2}\Bigg[\overline{\calZ_\RBIM(e,J)}+(-1)^j \, \overline{\calZ_\RBIM(e+\ell_1,J)}\Bigg]
\end{align}.
We may use Eq. \eqref{eq_Zising_vs_Zrbim} to substitute $\overline{\calZ_\RBIM(e,J)}$ with $\calZ_\Ising(\tilde J)$. However, first note that when on the Nishimori line (see Eq. \eqref{eq_nishimori}), one has $\tanh(J)=(1-2q)$. Then, by using Eq. \eqref{eq_q_vs_beta_apx} and Eq. \eqref{eq_Jtilde_vs_J}, one can find $\tilde{J}$ in terms of $\beta$. After few steps of algebra, we find that $\tilde{J}=\tilde{\beta}$ which is defined in Eq. \eqref{eq_betatilde_vs_beta}. Therefore, by substituting $\overline{\calZ_\RBIM(e,J)}$ with $\calZ_\Ising(\tilde \beta)$, and then using Eq. \eqref{eq_Z0_vs_Zising} and Eq. \eqref{eq_Z1_vs_Zising} to write the resulting expression in terms of $\calZ_j$ one finds,
\begin{align}
    &\sum_\bmm (P([e_\bmm])+(-1)^j P([e_\bmm+\ell_1]))^2=\nonumber\\
    &=\frac{[q(1-q)]^{N/2}}{2}\times  \Bigg[\frac{\cosh(J)}{\cosh(\tilde{\beta})}\Bigg]^{N}\times \frac{2 e^{N\tilde{\beta}}}{\cosh(\beta)^N}\calZ_j\\
    &=e^{-N\beta}\calZ_j=\calN_j,
\end{align}
where from second line to third line we have used Eq. \eqref{eq_q_vs_beta_apx}, Eq. \eqref{eq_nishimori} and Eq. \eqref{eq_betatilde_vs_beta} to write $q$, $J$ and $\tilde{\beta}$ in terms of $\beta$.

\section{Syndrome correlations in the perturbed toric code}\label{apx_synd_corr}
This section contains the derivation of Eq. \eqref{eq_synd_corr}. Let state $\ket{\psi}$ denote a state in the ground state subspace of $H(\beta)$. Since $m_s$ and $m_{s'}$ are measurement outcomes of $A_s$ and $A_{s'}$ operators, the connected correlation function $C$ can be written as,
\begin{align}
    C(s,s')=\expval{A_s A_{s'}}-\expval{A_s}\expval{A_{s'}},
\end{align}
where $\expval{}$ is the expectation value with respect to $\ket{\psi}$. 
As noted in Ref.\cite{castelnovo2008quantum}, the ground states of the perturbed toric code are annihilated by $Q_s=A_s-\exp(-\beta\sum_{j\in s}Z_j)$ for any star $s$ in lattice $\Lambda$, and thus $A_s \ket{\varphi_0(\beta)}=\exp(-\beta\sum_{j\in s}Z_j )\ket{\varphi_0(\beta)}$. Therefore,
\begin{align}
    C(s,s')=&\expval{e^{-\beta\sum_{j\in s,s'}Z_j }}\nonumber\\
    &-\expval{e^{-\beta\sum_{j\in s}Z_j }}\expval{e^{-\beta\sum_{j\in s'}Z_j}}.
\end{align}
To compute the expectation values, we follow similar reasoning as in Appendix \ref{apx_calZ_derivation}. First, assume $\ket{\psi}$ is  the $\ket{\varphi_0(\beta)}$ state. Then,
\begin{align}
    &\expval{e^{-\beta\sum_{j\in s}Z_j }}_0\nonumber\\
    &=\frac{1}{\calZ_0}\expval{e^{-\beta\sum_{j\in s}Z_j}e^{\beta\sum_{j}Z_j}}{\varphi_0(0)}\\
    &=\overline{\exp(-\beta \sigma_s\sum_{\langle \mu, s\rangle}\sigma_\mu )},
\end{align}
where the over line is the average with respect to the 2D Ising model, and $\mu$ sums over nearest neighbors of $s$. One would find a similar expression for the case when $\ket{\psi}=\ket{\varphi_1(\beta)}$, but the average is taken with respect to the Ising model with a domain-wall along the support of $\overline{X}$. More generally, when $\ket{\psi}=\alpha_0\ket{\varphi_0(\beta)}+\alpha_1\ket{\varphi_1(\beta)}$, one has $\expval{A_s}=|\alpha_0|^2 \expval{A_s}_0+|\alpha_1|^2 \expval{A_s}_1$, since $[A_s,\overline{Z}]=0$, where the subscripts $1,2$ indicate the expectation values is taken with respect to $\ket{\varphi_{1,2}(\beta)}$. When $\beta<\beta_c$, the system has a finite correlation length with no long range order. Moreover, note that we can always choose the support of $\overline{X}$ to be far from $s$ and $s'$, so the domain wall in the Ising model can be chosen to be $O(L)$ distance apart from $s$ and $s'$. Thus we expect $\expval{A_s}_1=\expval{A_s}_2$, up to errors which are suppressed in the system size. Therefore, if we define $\epsilon_s=-\sigma_s\sum_{\langle \mu, s\rangle}\sigma_\mu$ to denote the local energy at site $s$, we find that for any general state in the ground state subspace of $H(\beta)$ we have,
\begin{align}
    &C(s,s')=\overline{\exp(\beta \epsilon_s)\exp(\beta\epsilon_{s'})}-\overline{\exp(\beta \epsilon_s)}\times \overline{\exp(\beta \epsilon_{s'})},
\end{align}
up to errors which vanish in the thermodynamic limit, where the average is taken with respect to the 2D classical Ising model.

\section{Berry connection inside the gapped phase}\label{apx_closed_path_holonomy}
In this section we argue that under physically motivated assumptions the Berry connection vanishes in a neighborhood of the fixed point in the gapped phase. 
Let $\ket{\varphi_i(\blambda)}$ denote degenerate ground states at $\blambda$.
Importantly, we assume that $\ket{\varphi_i(\blambda)}$ is continuous functions of $\blambda$ in a neighborhood $\mathcal{A}$ around $\blambda=0$. Our goal is to show that $\mel{\varphi_j(\blambda)}{\partial_\blambda}{\varphi_i(\blambda)}$ vanishes in $\mathcal{A}$ in the thermodynamic limit.

Let $\delta \blambda=(\delta \lambda_1, \delta \lambda_2,,\cdots,\delta \lambda_n)$ denote a small change in parameter $\blambda$. We further assume that $\ket{\varphi_i(\blambda+\delta \blambda)}$ is related to $\ket{\varphi_i(\blambda)}$ via a unitary $u_{\delta \blambda}=1+i\sum_k \delta \lambda_k D_k + O(\delta \lambda^2)$, where $D_k$ is a sum of $O(N)$ local Hermitian operators with bounded norm. This assumption is justified by the fact that the states in a gapped phase are related to each other by a constant time local unitary evolution. Therefore,
\begin{align}
    \ket{\varphi_i(\blambda+\delta \blambda)}=&\ket{\varphi_i(\blambda)}+i\sum_{k}\delta \lambda_k D_k  \ket{\varphi_i(\blambda)}+O(\delta \lambda^2).
\end{align}
This shows that the components of the Berry connection are given as,
\begin{align}
    A_{ij;k}(\blambda)&=\mel{\varphi_j(\blambda)}{\partial_{\lambda_k}}{\varphi_i(\blambda)}\nonumber=i\mel{\varphi_j(\blambda)}{D_k}{\varphi_i(\blambda)}\\
    &=d_k(\blambda)\delta_{ij}+O(\exp(-L)).
\end{align}
To arrive at the second line we used the local indistinguishability of the ground states inside the gapped phase. This shows that the quantum holonomy $\mathcal{U}_\gamma$ in Eq. \eqref{eq_closed_path_holonomy} is just a global phase.

\section{Repetition code standard decoder and the phase diagram of the rotated Ising Model}\label{apx_decoding_rotated_ising_model}
In this section we show that the standard decoder of the repetition code (described in Section \ref{sec_recovery_map}) can be used to successfully recover the code states which were encoded in the ground state subspace of the Ising model $H_0=\sum_{i=1}^{N}Z_i Z_{i+1}$ but evolved adiabatically to the ground state subspace of $H_\theta$, with
\begin{align}
    H_\theta=U_\theta H_0 U^\dagger_\theta, \qquad U_\theta=\prod_{i=1}^N e^{-i\frac{\theta}{2}X_i},
\end{align}
for $-\pi/2 < \theta <\pi/2$.

If the initial state has been
\begin{align}
    \ket{\psi_0}=\alpha \ket{\ghz_+}+\beta\ket{\ghz_-},
\end{align}
the adiabatic evolution would transform it into,
\begin{align}
    \ket{\psi_\theta}=\alpha U_\theta \ket{\ghz_+}+\beta U_\theta \ket{\ghz_+},
\end{align}
up to a global phase. Imagine one measures all $Z_i Z_{i+1}$ stabilizers on this state, finding the syndrome $\bmm$, and hence inferring error $e_\bmm$ (see Section \ref{sec_recovery_map}). The probability of seeing syndrom $\bmm$ is given by,
\begin{align}
    \expval{\Pi_\bmm}{\psi_\theta}=&|\alpha|^2 \expval{U^\dagger_\theta\Pi_\bmm U_\theta}{\ghz_+}\nonumber \\
    &+|\beta|^2 \expval{U^\dagger_\theta\Pi_\bmm U_\theta}{\ghz_-}.
\end{align}
To compute $\expval{U^\dagger_\theta\Pi_\bmm U_\theta}{\ghz_\pm}$, we note that $U_\theta=\Pi_j \big[\cos(\frac{\theta}{2})+i \sin(\frac{\theta}{2}) X_j\big]$. Assuming $|e_\bmm|=k$, we have,
\begin{align}
    &\expval{U^\dagger_\theta\Pi_\bmm U_\theta}{\ghz_\pm}=\nonumber\\
    &\Bigg| \cos(\theta/2)^{N-k}(i\sin(\theta/2))^k\pm \cos(\theta/2)^{k}(i\sin(\theta/2))^{N-k}\Bigg|^2\nonumber\\
    =&\cos(\theta/2)^{2(N-k)}\sin(\theta/2)^{2k}\times \Big|1\pm (i\tan(\theta/2))^{N-2k}\Big|^2,
\end{align}
Note that $k/N\le 1/2$. Let us focus on $k/N<1/2 -\delta$ for some positive constant $\delta>0$, so we may drop the $\tan(\theta/2)$ term inside the absolute value for large $N$. In that case, the probability that the inferred error has support on $k$ qubits would simplify to,
\begin{align}
    p_k={ N \choose k}(\sin(\theta/2)^2)^k(\cos(\theta/2)^2)^{N-k}
\end{align}
which is a binomial distribution with mean,
\begin{align}\label{eq_apx_peak_of_k}
    k/N=\sin(\theta/2)^2
\end{align}.
For large $N$, the standard deviation of $k$ grows as $\sqrt{N}$, hence $k/N$ would be peaked around its mean with a width that goes to zero as $1/\sqrt{N}$. This justifies the assumption $k/N<1/2-\delta$ (Note that $-\pi/2<\theta<\pi/2$). Moreover, the above analysis also shows that after measuring the stabilizers and seeing syndrome $\bmm$, the state has been collapsed into,
\begin{align}
    \alpha e_\bmm \ket{\ghz_+}+\beta e_\bmm \ket{\ghz_-}+O(\tan(\theta/2)^{N-2k})
\end{align}
up to normalization. The error term is what results in the discrepancy between the recovered state and the original encoded state. However, since the probability distribution is peaked around $k/N=\sin(\theta/2)^2<1/2$, the typical error term vanishes exponentially in system size. Hence the average fidelity of the recovered state approaches $1$ in the thermodynamic limit.

\section{Decoding Adiabatic Noise in a $\mathbb{Z}_{2}\times \mathbb{Z}_{2}$ SPT State}\label{apx_clustermodel}
Consider the Hamiltonian for a one-dimensional cluster state in a transverse field, 
\begin{align}\label{eq_cluster_hamiltonian}
    H_c(g)=-\sum_{i=2}^{N-1}Z_{i-1}X_iZ_{i+1}-g\sum_i X_i,
\end{align}
on a chain of $N$ qubits, where $N$ is assumed to be even. For $b<1$, the system is in the SPT phase protected by the $\mathbb{Z}_2\times \mathbb{Z}_2$ symmetry generated by,
\begin{align}
    G_o=\prod_{i}X_{2i+1},\quad G_e=\prod_{i}X_{2i}.
\end{align}
On an open chain the ground state is four-fold degenerate, corresponding to two logical qubits. At $g=0$, the codespace is stabilized by all terms in the Hamiltonian, i.e. $Z_{i-1}X_iZ_{i+1}$ for $i=2,\cdots,N-1$.  The logical operators at $g=0$ thus could be taken to be,
\begin{align}
    &\overline{Z}_1=Z_1,\quad \overline{X}_1=G_o=\prod_{i}X_{2i+1},\\
    &\overline{Z}_2=Z_N,\quad \overline{X}_2=G_e=\prod_{i}X_{2i}.
\end{align}
Note that $Z_1 Z_{N-1}X_N$ is a non-trivial \textit{symmetric} logical operator, which is equal to $\overline{Z}_1 \overline{X}_2$ up to multiplication by stabilizers of the code. It means that even if one restricts to symmetric errors, the code distance is of order $\mathcal{O}(1)$. However, such symmetric errors are geometrically non-local and they are exponentially suppressed in any locally symmetric noise model, leading to the possibility of a finite error-threshold. The standard decoder for the cluster model is basically two copies of the standard decoder for the repetition code, one for even lattice sites and the other for the odd lattice sites. 

\begin{figure}
    \centering
    \includegraphics[width=\columnwidth]{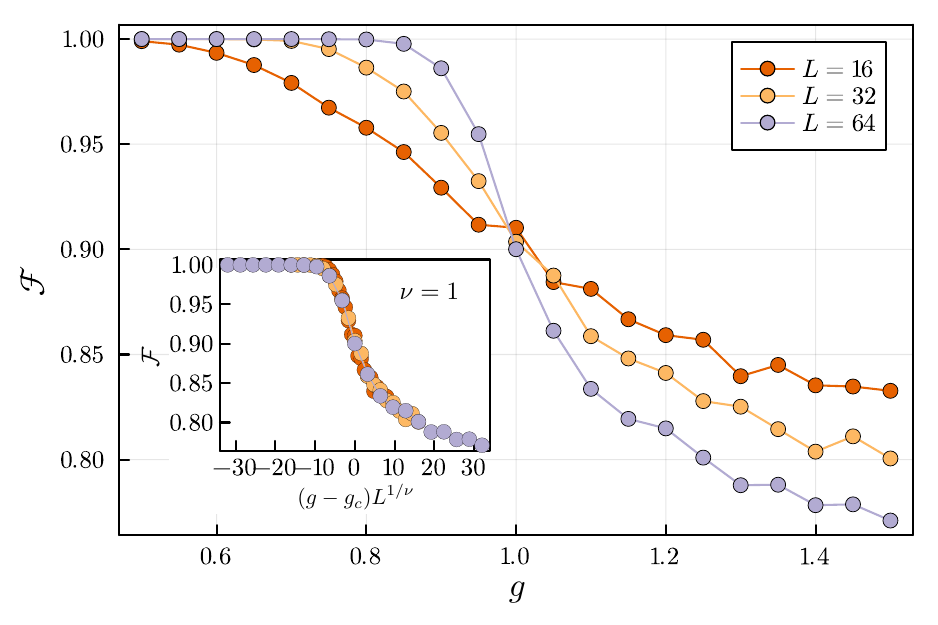}
    \caption{Fidelity of the recovered state with respect to the original encoded state (equal supper position of the four ground states  of  $H_c(0)$ with definite symmetries on an open chain) as a function of $g$. The inset shows the data collapse of the same data points, using $g_c=1$ and $\nu=1$.}
    \label{fig_F_ZXZ}
\end{figure}

Since $[H_c(g),G_e]=[H_c(g),G_o]=0$, $H_c(g)$ can be turned into a block diagonal form, where each block correspond to specific charges under $G_e$ and $G_o$. Let $\ket{\varphi_{\pm,\pm}(g)}$, denote the four ground sates in the four charge sectors, where the first (second) sign denotes the symmetry charge under $G_o$ ($G_e$) symmetry. The global phase of $\ket{\varphi_{\pm,\pm}(g)}$ is fixed by demanding $\expval{\partial_g}{\varphi_{\pm,\pm}(g)}$ to vanish. Similar to the non-integrable model studied in Appendix \ref{apx_nonintg_tfim}, we satisfy this condition by fixing $\braket{\varphi_{\pm,\pm}(0)}{\varphi_{\pm,\pm}(g)}$ to be real. Fig.\ref{fig_F_ZXZ} shows the average fidelity of the recovered state, and the originally encoded state which we have taken to be $\ket{\psi_0}=\frac{1}{2}\sum_{c_1=\pm,c_2=\pm}\ket{\varphi_{c_1,c_2}(0)}$. The inset shows  the corresponding data collapse with $\nu=1$ and $g_c=1$. As can be seen from the plot, perfect recovery is possible via the standard decoder throughout the entire SPT phase. 

The success of the standard decoder can be understood b following the same line of argument as for the repetition code. Note that the SPT phase is characterized by non-vanishing of string order parameters $\expval{s_{i,j}}$ when $|i-j|\gg 1$. Moreover, the string order parameters $s_{i,j}$ are basically the product of the stabilizers $Z_{i-1}X_iZ_{i+1}$ from $i$ to $j$. This is analogous to $\expval{Z_i Z_j}$ order parameter in the ferromagnetic phase and the fact that $Z_i Z_j$ is just the product of the  stabilizers of the repetition code. As such, non-zero expectation value of string order parameters alongside the indistinguishably of the ground states throughout the SPT phase ensures that the  syndrome data would not reveal any information about the logical sector and the decoding step would not implement any erroneous logical operation. We expect the same result to hold for any 1D SPT phase whenever the symmetry group is finite and Abelian and the SPT phase corresponds to a maximally non-commutative representation of it, since each step of the above argument can be generalized in that case. 

\end{document}